\documentclass[%
 preprint, 
 nofootinbib,
 amsmath,amssymb,
 aps, physrev,
]{revtex4-2}

\usepackage{graphicx}
\usepackage{dcolumn}
\usepackage{bm}
\usepackage{amsmath}
\begin{document}

\title{\textbf{Beam tracing and profile evolution for localised beams in inhomogeneous plasmas} 
}%

\author{Lewin B.S Marsh}
\email{lewin.marsh@gmail.com}
\affiliation{
  University of Oxford, OX1 1NY, UK
}
\author{Felix I. Parra}
\affiliation{
  Princeton Plasma Physics Laboratory, Princeton, NJ 08540, USA
}
\author{Valerian Hongjie Hall-Chen}
\affiliation{
  Future Energy Acceleration and Translation Programme, Agency of Science,
Technology and Research (A*STAR), Singapore 138632, Singapore
}
\author{Juan Ruiz Ruiz}
\affiliation{
  Rudolf Peierls Centre for Theoretical Physics, University of Oxford, OX1 3NP, UK
}

\date{\today}

\begin{abstract}
We derive the beam tracing and profile evolution for the propagation of any localised beam with arbitrary profile through an inhomogeneous cold plasma. We recover standard Gaussian beam-tracing, with an additional PDE describing the evolution of the beam's profile as it propagates through the plasma. We then solve for generic families of solutions to the PDE using ladder operators, which can be chosen to reduce to Gauss-Hermite beams in homogeneous media. We importantly obtain an exact expression for the resulting beam profile, demonstrating that Hermite modes will generally evolve into a superposition of different modes during propagation through inhomogeneous plasmas, contrary to prior work on the subject. Importantly, this theory provides us with a lightweight numerical approach for modelling microwave beam propagation through a plasma, which is easily implemented on top of existing Gaussian beam tracing techniques.
\end{abstract}

\maketitle

\section{Introduction}
The propagation of electromagnetic beams through plasmas has long been an important aspect of nuclear fusion research. The ability to model and interpret the evolution of a beam's profile and power deposition has proven to be important parts of our plasma control toolbox, via heating, current drive\cite{Erckmann1994, wolf2018, sysoeva2015}, and even suppression of MHD instabilities\cite{LaHaye2009}. On top of this, microwave diagnostic techniques are versatile tools for real-time plasma diagnostics. With appropriate engineering, they can be deployed in environments with high neutron flux\cite{Goncalves2023engineering}, as well as provide a broad range of measurements -- we diagnose electron temperature via electron-cyclotron emission(ECE)\cite{bornatici1983ece, classen2010ece, liu2018ece}, electron density via Doppler reflectometry\cite{Hillesheim2009, Conway2004, gusakov2004, Lin2024reflectometry}, and line-integrated magnetic information via polarimetry\cite{smith2008polarimetry,zhu2015polarimetry,chen2018polarimetry}. Notably, techniques like Doppler backscattering(DBS)\cite{HallChen2022,HallChen2022b,HallChen2024,liang2025,yeoh2025,RuizRuiz2025, Hennequin2006} and cross-polarisation scattering(CPS)\cite{stepanov1998,colas1996} further enable studies of electromagnetic turbulence and MHD activity with higher temporal and spatial resolution. On top of this, many of the aforementioned systems are capable of measuring even more plasma properties with modern advancements\cite{Kohn2025review}. Given the incredible robustness of these microwave techniques, accurate modelling of microwave interactions and measurements is essential.\\
\\
Historically, ray-tracing codes\cite{smirnov2001genray, Esterkin1996, xie2022boray} are typically sufficient for understanding the propagation of a beam through a plasma, by performing ray tracing across multiple(possibly coupled) rays to model how the beam profile might evolve. Beam tracing offers faster and more accurate simulations by modelling the evolution of the profile of the beam itself. This allows us to model diffraction effects with greater accuracy compared to a simplistic ray description of a beam. TORBEAM\cite{Poli2001, Poli2018}, which was built upon theoretical work by Pereversev\cite{Pere1998}, was one of the first codes developed for modelling beam evolution using beam tracing. It has certainly set the groundwork for synthetic diagnostic and heating techniques, while newer codes like Scotty\cite{HallChen2022,Wikarta2025}, WKBeam\cite{weber2015, Snicker2018} and PARADE\cite{PARADE2021} have since been developed and further optimised for their specific use cases, instead of being one-size-fits-all codes like TORBEAM. \\
\\
Given the prevalence and pertinence of synthetic beam tracing applications in plasma diagnostics and control, we revisit the propagation of an arbitrary beam profile in an anisotropic plasma. Whilst the propagation of the lowest order Gauss-Hermite mode(a Gaussian beam) has been well understood, the higher order modes are still not that well understood. Pereversev's early theoretical work on the subject utilised asymptotic expansion of a small parameter \(\sqrt{\kappa}\), similar to the parameter \(\frac{\lambda}{W}\) we will be using in this work. The issue with this approach is that these orderings break down near a caustic, making them unreliable near caustics. On the other hand, metaplectic geometrical optics(MGO) is the most modern approach currently in use. It generalizes beam tracing whilst remaining valid through caustics and mode-conversion regions\cite{lopez2022metaplectic,lopez2024,donnelly2021steepest,hojlund2024metademo}. In the paraxial limit, MGO reduces to standard beam tracing. Despite this, we focus on solving for the propagation of electromagnetic beams in cold plasmas exactly using the asymptotic expansion approach. This approach is mathematically more tractable than MGO, and still provides us with important insights away from caustics. Our goal is not only to model the propagation of microwave beams in general, but to set the theoretical groundwork for improved accuracy in signal analysis, by modelling the evolution of non-Gaussian imperfections in the antenna pattern.\\
\\
Whilst it may seem that this would offer nothing new that Pereversev's approach did not already show, we surprisingly discover that beams that start out as Gauss-Hermite beams may evolve into some other kind of profile, which we ascertained exactly to be a linear combination of Hermite modes in a rotated basis. This is in contrast to Pereversev's results, suggesting that this approach has not yet been well-understood and deserves more attention. We discuss this potential discrepancy in greater detail in section \ref{pereversev_discussion}. Apart from simple application to a Gauss-Hermite basis, our work has the advantage of providing immense freedom to generate many different bases of solutions, which allows us to model the evolution of each basis of solutions exactly. This should allow one to choose a basis most suited for their problem at hand, for example, when choosing between `elegant' and `traditional' Hermite modes in Gaussian wave optics\cite{pampaloni2004}. \\
\\
We first derive beam-tracing for any beam using asymptotic expansion up to the second order in an inhomogeneous cold plasma. We make a distinction between the Gaussian beam envelope and the beam profile, which is some function attenuated by said envelope. This is purely a mathematical choice. We demonstrate in the case of a Gaussian beam that the exact distinction of `envelope' and `profile' do not affect the consistency of our results, as one would expect. We discover that the evolution of our beam is governed by the same set of beam-tracing equations for a pure Gaussian beam envelope as in prior, separate work by Pereversev and Hall-Chen et al.\cite{HallChen2022}, with an added second-order partial differential equation governing the evolution of the beam profile. The profile evolution equation we obtain demonstrates that Hermite polynomial beam profiles are generally not preserved in an inhomogeneous plasmas, although a constant profile expectedly remains unchanged. \\
\\
We proceed to solve the PDE in general, using ladder operators to generate two infinite, biorthogonal families of solutions. In particular, we focus on constructing ladder operators that reduce to Gauss-Hermite polynomials at the plasma-vacuum boundary. We confirm that Gauss-Hermite polynomials are solutions in homogeneous, isotropic media, in agreement with the conventional wisdom. In general, the solutions do not have to be Hermite polynomials at the plasma boundary, and there are many other solutions that could be utilised. 
Finally, we incorporate weak dissipative effects for modelling microwave heating of the plasma.\\
\\
We finally discuss potential future implementation and  applications of our work in microwave diagnostics.

\section{Beam Tracing Equations}
In this section, we lay out the derivation of our beam-tracing equations, utilising perturbation theory. To perform the expansion, we will order that \({\lambda}/{W}\sim{W}/{L}\ll1\). \(\lambda\) refers to the wavelength of our beam, \textit{W} refers to the width of our beam, and \textit{L} refers to the typical length scale of inhomogeneity in our plasma.\\
\\
Our goal is to find the beam solution to the Helmholtz equation in a plasma. Repeated indices will be summed over whenever present in this manuscript. Occasionally, Einstein notation is utilised when it presents greater simplicity than tensor notation. We are ultimately interested in finding solutions to the Helmholtz equation for an electromagnetic wave, which is given by the following in a plasma:
\begin{equation}
    \frac{c^2}{\Omega^2}\boldsymbol{\nabla}\times(\boldsymbol{\nabla}\times\boldsymbol{E})=\boldsymbol{\epsilon}\cdot\boldsymbol{E}.
\end{equation}
\(\boldsymbol{E}\) refers to the components of the electric field at any point without the time factor \(\exp(-\mathrm{i}\omega t)\), and \(\boldsymbol{\epsilon}\) is the cold plasma dielectric tensor. Note that we will be utilising the Euclidean metric for all summations, thus there is no difference between upper and lower indices, and they will be used as needed to keep the notation readable.\\
\\
In order to give a beam solution to the Helmholtz equation, we need to define the beam ansatz.
\subsection{Beam Ansatz}
The beam ansatz we define only requires our beam to be localised in space, but it can have any cross section within that localised envelope. The coordinates of our central ray are given by \(\bm{q}(\tau)\) where \(\tau\) simply parametrises the path. The ray velocity at every point is given by \(\bm{g}=g\hat{\bm{g}}={\mathrm{d}\bm{q}}/{\mathrm{d}\tau}\). \(\bm{w}\) is the width perpendicular to said ray velocity vector at any point. We thus adopt the following beam ansatz:
\begin{equation}\label{eqn:ansatz}
\begin{split}
    \bm{E}(\bm{r})&=\bm{A}\left(\tau,w_x,w_y\right)\exp\left(\mathrm{i}\psi\right),\\
    \psi&=s(\tau)+\bm{K}_w\cdot\bm{w}+\frac{1}{2}\bm{w}\cdot\bm{\Psi}_w\cdot\bm{w},
\end{split}
\end{equation}
where \(w_x\) and \(w_y\) are components of \(\bm{w}\) in the orthonormal basis \((\hat{\boldsymbol{g}},\hat{\boldsymbol{x}},\hat{\boldsymbol{y}})\)\footnote{We only require $(\hat{\boldsymbol{g}},\hat{\boldsymbol{x}},\hat{\boldsymbol{y}})$ to be an orthonormal basis, and impose no other restrictions on the orientations of $(\hat{\boldsymbol{x}},\hat{\boldsymbol{y}})$. In Hall-Chen et. al\cite{HallChen2022}, an orientation is explicitly chosen, though such a choice was not strictly necessary for the beam tracing derivation.}, where we have freedom in the basis orthogonal to \(\hat{\boldsymbol{g}}\). \(\bm{\Psi}_w\) is a complex tensor, where the imaginary part encodes the beam envelope while the real part encodes the curvature of the beam. \(s(\tau)=\int_0^\tau K_g(\tau')g\mathrm{d}\tau'\) is a phase term included such that together with \(\bm{K_w}\), we have a well defined wavevector \(\bm{K}\) for our beam. In essence, \(s(\tau)\) encodes the change in phase due to propagation along the group velocity. Although we could always absorb the envelope and phase into the amplitude function \(\bm{A}\), keeping it separate yields simpler and more illuminating mathematics. These quantities we have defined have the following orderings: \(\bm{K}\sim{1}/{\lambda}\) and \(\bm{\Psi}_w\sim{1}/{W^2}\). Thus, \(s\sim{L}/{\lambda}\) and \(\bm{K}_w\cdot\bm{w}\sim{W}/{\lambda}\).\\
\\
As we will be performing perturbation theory later on, we immediately present the chosen expansion for our amplitude function:
\begin{equation}\label{eqn:amplitude_ansatz}
    \bm{A}\left(\tau,w_x,w_y\right)=\mathcal{P}\left(\tau,w_x,w_y\right)A^{(0)}(\tau)\hat{\bm{e}}^{(0)}(\tau)+\bm{A}^{(1)}\left(\tau,w_x,w_y\right)+\bm{A}^{(2)}\left(\tau,w_x,w_y\right)+... 
\end{equation}
The superscripts indicate the order in \(\frac{\lambda}{W}\) of the associated quantity. To lowest order, we have chosen to decompose our amplitude into a profile function \(\mathcal{P}\) as well as an amplitude \(A^{(0)}\). This choice is primarily for compatibility with previous results and simplicity, but it also makes classifying beams with different profiles easier later on. It is important that our chosen profile must have weaker growth than our Gaussian beam envelope at all times, in order to be a localised beam.\\
\\
Finally, in order to substitute this ansatz into the Helmholtz equation, we need to transform our gradient operator into our beam-centred coordinates. If we choose some orientation for our beam width, such that \(\bm{w}=w_x\hat{\bm{x}}+w_y\hat{\bm{y}}\) it is given by:
\begin{equation}
    \bm{\nabla}=\bm{\nabla}\tau\frac{\partial}{\partial\tau}+\bm{\nabla}w_x\frac{\partial}{\partial w_x}+\bm{\nabla}w_y\frac{\partial}{\partial w_y}.
\end{equation}
As derived in previous work\cite{HallChen2022}, these quantities are given by:
\begin{equation}
\begin{split}
    \bm{\nabla}\tau&=\frac{\hat{\bm{g}}}{g(1-\bm{\kappa}\cdot\bm{w})},\\
    \bm{\nabla}w_x&=\hat{\bm{x}}+\frac{w_y\hat{\bm{g}}}{g(1-\bm{\kappa}\cdot\bm{w})}\frac{\mathrm{d}\hat{\bm{x}}}{\mathrm{d}\tau}\cdot\hat{\bm{y}},\\
    \bm{\nabla}w_y&=\hat{\bm{y}}+\frac{w_x\hat{\bm{g}}}{g(1-\bm{\kappa}\cdot\bm{w})}\frac{\mathrm{d}\hat{\bm{y}}}{\mathrm{d}\tau}\cdot\hat{\bm{x}},\\
\end{split}
\end{equation}
where \(\bm{\kappa}\) is the ray curvature, given by \(\bm{\kappa}=({1}/{g})\,({\mathrm{d}\hat{\bm{g}}}/{\mathrm{d}\tau})\sim{1}/{L}\). In general, all derivatives along the ray length would be on the length scale of inhomogeneity of our plasma. Using this our, gradient operator thus has the following form:
\begin{equation}
    \boldsymbol{\nabla}=\underbrace{\hat{\bm{x}}\frac{\partial}{\partial w_x}+\hat{\bm{y}}\frac{\partial}{\partial w_y}}_{\boldsymbol{\nabla}_w}+\frac{\hat{\bm{g}}}{g(1-\bm{\kappa}\cdot\bm{w})}\left(\frac{\partial}{\partial\tau}-\frac{\mathrm{d}\bm{w}}{\mathrm{d}\tau}\cdot\boldsymbol{\nabla}_w\right)
\end{equation}
We now substitute our ansatz in eq. \ref{eqn:ansatz} into the Helmholtz equation. We will use \(\bm{E}=\bm{A}\exp(\mathrm{i}\psi)\). The right hand side of the Helmholtz equation can be expanded about the center of the ray, at any point along the ray:
\begin{equation}
    \bm{\epsilon}(\bm{r})=\bm{\epsilon}(\bm{q}(\tau))+\boldsymbol{w}\cdot\boldsymbol{\nabla}\bm{\epsilon}(\bm{q}(\tau))+\frac{1}{2}\boldsymbol{ww}:\boldsymbol{\nabla}\boldsymbol{\nabla}\bm{\epsilon}(\bm{q}(\tau))+...
\end{equation}
On the left hand side of the Helmholtz equation, the derivatives of the electric field are given by:
\begin{equation}
\begin{split}
\partial_j\partial_l(\boldsymbol{E})&=(\partial_j\partial_l \boldsymbol{A}+\mathrm{i}\left(\partial_j\psi\partial_l \boldsymbol{A}+\partial_l\psi\partial_j \boldsymbol{A}\right)\\
&-\partial_j\psi\partial_l\psi \boldsymbol{A}+\mathrm{i}\partial_j\partial_l\psi \boldsymbol{A})\exp(\mathrm{i}\psi).\\
\end{split}
\end{equation}
We thus require all these various derivatives, along with their expansions and orderings. For the derivatives of the phase $\psi$, using the fact that:
\begin{equation}
    \left(\frac{\mathrm{d}\bm{K}_w}{\mathrm{d}\tau}\right)_w=\left(\frac{\mathrm{d}K_x}{\mathrm{d}\tau}+K_y\frac{\mathrm{d}\hat{\bm{y}}}{\mathrm{d}\tau}\cdot \hat{\bm{x}}\right)\hat{\bm{x}}+\left(\frac{\mathrm{d}K_y}{\mathrm{d}\tau}+K_x\frac{\mathrm{d}\hat{\bm{x}}}{\mathrm{d}\tau}\cdot \hat{\bm{y}}\right)\hat{\bm{y}},
\end{equation}
and:
\begin{equation}
\begin{split}
    \left(\frac{\mathrm{d}\bm{\Psi}_w}{\mathrm{d}\tau}\right)_w&=\left(\frac{\mathrm{d}\Psi_{xx}}{\mathrm{d}\tau}+2\Psi_{xy}\frac{\mathrm{d}\hat{\bm{y}}}{\mathrm{d}\tau}\cdot \hat{\bm{x}}\right)\hat{\bm{x}}\hat{\bm{x}}\\
    &+\left(\frac{\mathrm{d}\Psi_{xy}}{\mathrm{d}\tau}+\Psi_{xx}\frac{\mathrm{d}\hat{\bm{x}}}{\mathrm{d}\tau}\cdot \hat{\bm{y}}+\Psi_{yy}\frac{\mathrm{d}\hat{\bm{y}}}{\mathrm{d}\tau}\cdot \hat{\bm{x}}\right)(\hat{\bm{x}}\hat{\bm{y}}+\hat{\bm{y}}\hat{\bm{x}})\\
    &+\left(\frac{\mathrm{d}\Psi_{yy}}{\mathrm{d}\tau}+2\Psi_{xy}\frac{\mathrm{d}\hat{\bm{x}}}{\mathrm{d}\tau}\cdot \hat{\bm{y}}\right)\hat{\bm{y}}\hat{\bm{y}},
\end{split}
\end{equation}
we find that:
\begin{equation}
    \bm{\nabla}\psi=\bm{K}+\frac{\hat{\bm{g}}}{g(1-\bm{\kappa}\cdot\bm{w})}\left(\frac{\mathrm{d}\bm{K}}{\mathrm{d}\tau}\cdot\bm{w}\right)+\bm{\Psi}_w\cdot\bm{w}+\frac{\hat{\bm{g}}}{g(1-\bm{\kappa}\cdot\bm{w})}\left(\frac{1}{2}\bm{w}\cdot\frac{\mathrm{d}\bm{\Psi}_w}{\mathrm{d}\tau}\cdot\bm{w}\right).
\end{equation}
Thus, to get the various orders of \(\bm{\nabla}\psi\), we simply have to expand \((1-\bm{\kappa}\cdot\bm{w})^{-1}=1+\bm{\kappa}\cdot\bm{w}+(\bm{\kappa}\cdot\bm{w})^2+...\), as each term is an order \({\lambda}/{W}\) lower than the previous. As for the second order derivatives, we will only require \((\bm{\nabla}\bm{\nabla}\psi)^{(0)}\). We summarise all of the derivatives of the phase required in our derivation below:
\begin{equation}
\begin{split}
(\bm{\nabla}\psi)^{(0)}&=\bm{K},\\
(\bm{\nabla}\psi)^{(1)}&=\frac{\hat{\bm{g}}}{g}\left(\frac{\mathrm{d}\bm{K}}{\mathrm{d}\tau}\cdot\bm{w}\right)+\bm{\Psi}_w\cdot\bm{w},\\
&=\bm{\Psi}\cdot\bm{w},\\
(\bm{\nabla}\psi)^{(2)}&=(\bm{\kappa}\cdot\bm{w})\frac{\hat{\bm{g}}}{g}\left(\frac{\mathrm{d}\bm{K}}{\mathrm{d}\tau}\cdot\bm{w}\right)+\frac{\hat{\bm{g}}}{g}\left(\frac{1}{2}\bm{w}\cdot\frac{\mathrm{d}\bm{\Psi}_w}{\mathrm{d}\tau}\cdot\bm{w}\right),\\
(\bm{\nabla}\bm{\nabla}\psi)^{(0)}&=\bm{\Psi}_w+\frac{\hat{\bm{g}}}{g}\frac{\mathrm{d}\bm{K}}{\mathrm{d}\tau}+\left(\frac{\mathrm{d}\bm{K}}{\mathrm{d}\tau}\right)_w\frac{\hat{\bm{g}}}{g}=\bm{\Psi}\sim\frac{1}{W^2},\\
\Psi_{\mu\nu}&=\Psi_{\nu\mu}.\\
\end{split}
\end{equation}
We have defined a generalised form of \(\bm{\Psi}_w\) called \(\bm{\Psi}\), as the projection onto the width of said quantity will provide us with precisely \(\bm{\Psi}_w\). Next, we wish to obtain the ordering for the derivatives of our amplitude function. We once again summarise any necessary derivatives below:
\begin{equation}
\begin{split}
(\bm{\nabla}\boldsymbol{A})^{(0)}&=(\bm{\nabla}\mathcal{P})^{(0)}A^{(0)}\hat{\bm{e}},\\
(\bm{\nabla}\boldsymbol{A})^{(1)}&=(\bm{\nabla}\mathcal{P})^{(1)}A^{(0)}\hat{\bm{e}}+\mathcal{P}\left(\bm{\nabla}(A^{(0)}\hat{\bm{e}})^{(0)}+(\bm{\nabla}\boldsymbol{A}^{(1)})^{(0)}\right),\\
(\bm{\nabla}\bm{\nabla}\boldsymbol{A})^{(0)}&=(\bm{\nabla}\bm{\nabla}\mathcal{P})^{(0)}A^{(0)}\hat{\bm{e}}\sim\frac{A^{(0)}}{W^2}.\\
\end{split}
\end{equation}
One can notice that \(\bm{\nabla}\psi^{(n)}\sim\lambda^{-1}({\lambda}/{W})^n\) whilst \((\bm{\nabla}\boldsymbol{A})^{(n)}\sim({A^{(0)}}/{W})({\lambda}/{W})^n\). The derivatives of various orders of our profile function are given by:
\begin{equation}
\begin{split}
    (\bm{\nabla}\mathcal{P})^{(0)}&=\left(\hat{\bm{x}}\frac{\partial}{\partial w_x}+\hat{\bm{y}}\frac{\partial}{\partial w_y}\right)\mathcal{P}=\boldsymbol{\nabla}_w\mathcal{P},\\
    (\bm{\nabla}\mathcal{P})^{(1)}&=\left(\frac{\hat{\bm{g}}}{g}\frac{\partial}{\partial \tau}+\frac{w_y\hat{\bm{g}}}{g}\frac{\mathrm{d}\hat{\bm{x}}}{\mathrm{d}\tau}\cdot\hat{\bm{y}}\frac{\partial}{\partial w_x}+\frac{w_x\hat{\bm{g}}}{g}\frac{\mathrm{d}\hat{\bm{y}}}{\mathrm{d}\tau}\cdot\hat{\bm{x}}\frac{\partial}{\partial w_y}\right)\mathcal{P},\\
    &=\frac{\hat{\bm{g}}}{g}\left(\frac{\partial}{\partial\tau}-\frac{\mathrm{d}\bm{w}}{\mathrm{d}\tau}\cdot\boldsymbol{\nabla}_w\right)\mathcal{P},\\
    (\bm{\nabla}\bm{\nabla}\mathcal{P})^{(0)}&=\left(\hat{\bm{x}}\frac{\partial}{\partial w_x}+\hat{\bm{y}}\frac{\partial}{\partial w_y}\right)\left(\hat{\bm{x}}\frac{\partial}{\partial w_x}+\hat{\bm{y}}\frac{\partial}{\partial w_y}\right)\mathcal{P}=\boldsymbol{\nabla}_w\boldsymbol{\nabla}_w\mathcal{P}.
\end{split}
\end{equation}
We have ordered our terms such that the lowest order derivative of $\mathcal{P}$ with respect to the beam width is of the length scale $\sim W$, whilst the point-to-point variation in our profile must be of the typical length scale of the system $\sim L$. Explicitly:
\begin{equation}
    \frac{|(\bm{\nabla}\mathcal{P})^{(0)}|}{\mathcal{P}}\sim\frac{1}{W}\quad \mathrm{and}\quad\left|\frac{\partial\mathcal{P}}{\partial\tau}\right|\sim\frac{g}{L}{\mathcal{P}}.
\end{equation}
With all these prerequisites established, we can finally perform the expansions smoothly. We will have to expand up to second order in order to obtain the lowest order beam-tracing equations we require.

\subsection{The beam tracing equations}\label{beam-tracing-eqns}
For the sake of brevity, we go through the methodologically simple, but algebraically complicated derivation of the beam tracing equations in appendix \ref{beam_tracing_derivation}. All in all, we obtain 6 equations which not only model the propagation of our central ray, but completely describe the evolution of the envelope, amplitude, and profile of the beam. \\
\\
At 0-th order, the first and most basic equation is the dispersion relation, given by:
\begin{equation}
    \bm{D}\cdot\hat{\bm{e}}=H\hat{\bm{e}}=0,
\end{equation}
where:
\begin{equation}\label{eqn:dispersion relation}
    \bm{D}(\bm{K},\bm{r})=\frac{c^2}{\Omega^2}(\bm{KK}-K^2\bm{1})+\bm{\epsilon}.
\end{equation}
From this definition of the Hamiltonian using the dispersion relation, we can solve for the ray tracing equations, which we obtain from the 1-st order expansion. These tell us the trajectory of our central ray, along which we require the dispersion relation to be satisfied. These equations are:
\begin{equation}\label{eqn:ray-tracing}
\begin{split}
    \bm{\nabla}_{\boldsymbol{K}}H&=\frac{\mathrm{d}\bm{q}}{\mathrm{d}\tau},\\
    \bm{\nabla}H&=-\frac{\mathrm{d}\bm{K}}{\mathrm{d}\tau}.
\end{split}
\end{equation}
At second order, we obtain a second order PDE for the evolution of the beam, which is the eikonal equation. We firstly obtain the beam envelope evolution equation by simply defining it to be:
\begin{equation}\label{eqn:Beam_Tracing}
    \frac{\mathrm{d}\bm{\Psi}}{\mathrm{d}\tau}+\bm{\Psi}\cdot\bm{\nabla}_{\bm{K}}\bm{\nabla}_{\bm{K}}(H)\cdot\bm{\Psi}+\bm{\Psi}\cdot\bm{\nabla}_{\bm{K}}\bm{\nabla}(H)+\bm{\nabla}\bm{\nabla}_{\bm{K}}(H)\cdot\bm{\Psi}+\bm{\nabla}\bm{\nabla}(H)=0,
\end{equation}
Next are the amplitude evolution equations, where \(A^{(0)}=|A^{(0)}|\exp(\mathrm{i}\phi)\). We also define them such that we have the following equation for the magnitude of the complex amplitude:
\begin{equation}\label{eqn:Amplitude_Evolution_magnitude}
    |A^{(0)}|=C\frac{\mathrm{Det(}\Im(\bm{\Psi}))^\frac{1}{4}}{g^\frac{1}{2}},
\end{equation}
and the phase of the complex amplitude:
\begin{equation}\label{eqn:Amplitude_Evolution_Phase}
    \frac{\mathrm{d}\phi}{\mathrm{d}\tau}=\mathrm{i}\hat{\bm{e}}^*\cdot\frac{\mathrm{d}\hat{\bm{e}}}{\mathrm{d}\tau}+\frac{\mathrm{i}}{2}\left(\frac{\partial \hat{\bm{e}}^*}{\partial r_\mu}\cdot \bm{D}\cdot\frac{\partial \hat{\bm{e}}}{\partial K_\mu}-\frac{\partial \hat{\bm{e}}^*}{\partial K_\mu}\cdot \bm{D}\cdot\frac{\partial \hat{\bm{e}}}{\partial r_\mu}\right)-\frac{1}{2}\Im(\bm{\Psi}):\bm{\nabla}_{\bm{K}}\bm{\nabla}_{\bm{K}} H.
\end{equation}
We define the polarisation phase as:
\begin{equation}
    \frac{\mathrm{d}\phi_P}{\mathrm{d}\tau}=\mathrm{i}\hat{\bm{e}}^*\cdot\frac{\mathrm{d}\hat{\bm{e}}}{\mathrm{d}\tau}+\frac{\mathrm{i}}{2}\left(\frac{\partial \hat{\bm{e}}^*}{\partial r_\mu}\cdot \bm{D}\cdot\frac{\partial \hat{\bm{e}}}{\partial K_\mu}-\frac{\partial \hat{\bm{e}}^*}{\partial K_\mu}\cdot \bm{D}\cdot\frac{\partial \hat{\bm{e}}}{\partial r_\mu}\right),
\end{equation}
while the Gouy phase is given by:
\begin{equation}
    \frac{\mathrm{d}\phi_G}{\mathrm{d}\tau}=-\frac{1}{2}\Im(\bm{\Psi}):\bm{\nabla}_{\bm{K}}\bm{\nabla}_{\bm{K}} H.
\end{equation}
All these definitions reduce our second-order PDE into this much simpler equation defining the evolution of our beam profile:
\begin{equation}\label{eqn:Profile_Evolution}
    \frac{\partial\mathcal{P}}{\partial\tau}-\mathrm{i}\left(\frac{1}{2}\bm{\nabla}_{\bm{K}}\bm{\nabla}_{\bm{K}} H:\bm{\nabla}_w\bm{\nabla}_w\mathcal{P}\right)+\bm{w}\cdot\bm{T}_w\cdot \bm{\nabla}_w\mathcal{P}=0.
\end{equation}
where \(\bm{T}\) is given by:
\begin{equation}
\begin{split}
    \bm{T}(\tau)&=\left(\frac{\mathrm{d}\hat{\bm{y}}}{\mathrm{d}\tau}\cdot\hat{\bm{x}}\right)(\hat{\bm{x}}\hat{\bm{y}}-\hat{\bm{y}}\hat{\bm{x}})+\Big(\bm{\Psi}\cdot\bm{\nabla}_{\bm{K}}\bm{\nabla}_{\bm{K}} H+\bm{\nabla}\bm{\nabla}_{\bm{K}} H\Big),
\end{split}
\end{equation}
and \(\bm{T}_w\) is the projection onto the width, given by \(\bm{T}_w=(\bm{1}-\hat{\bm{g}}\hat{\bm{g}})\cdot\bm{T}\cdot(\bm{1}-\hat{\bm{g}}\hat{\bm{g}})\). This simple form of our second order PDE is why we chose to separate the profile and amplitude functions from this arbitrarily chosen `profile' function. This equation is similar to the Ornstein-Uhlenbeck equation\cite{Vatiwutipong2019}. Whilst this is a great simplification from the Helmholtz equation(a 3D non-linear PDE), solving it is not computationally efficient. \\
\\
All of these equations are ODEs which can be simply integrated over, with the exception of our profile evolution equation. While we have efficient methods for solving ODEs, solving a 2D linear PDE is still computationally more demanding. The primary purpose of our work is to be able to provide real-time beam analysis, thus possessing a faster approach would be desirable. In the subsequent section, we will do just that, by solving for generic families of solutions to eq. \ref{eqn:Profile_Evolution} via repeated application of ladder operators.

\section{Ladder operator solutions of profile evolution equation\label{Solving_Profile_Evolution}}
In this section, we present general operations that could yield the evolution of any initially confined beam profile via ladder operators\cite{Leen2016}. We are clearly interested solutions to eq. \ref{eqn:Profile_Evolution}, so we are finding eigenfunctions of the differential operator \(\hat{\mathcal{D}}\) given in index notation by:
\begin{equation}
    \hat{\mathcal{D}}=\frac{\partial}{\partial\tau}-\mathrm{i}\frac{1}{2}\frac{\partial}{\partial K_\mu} \frac{\partial}{\partial K_\nu} H\frac{\partial}{\partial w_\mu}\frac{\partial}{\partial w_\nu}+w_\mu T_{\mu\nu}\frac{\partial}{\partial w_\nu},
\end{equation}
where $\frac{\partial}{\partial w_\mu}:=\nabla_{w,\mu}$ such that:
\begin{equation}
    \hat{\mathcal{D}}f(\tau,w_x,w_y)=0.
\end{equation}
Although this approach gives us the freedom to model the evolution of a multitude of initial profiles, our end goal will be to construct solutions that reduce to Gauss-Hermite beams at the plasma-vacuum boundary. If we could do so, we can utilise the orthonormality of the Hermite polynomials to decompose any function into a sum of Gauss-Hermite modes, each of which we can evolve exactly. Furthermore, the Gauss-Hermite solutions have been studied before, allowing us to discuss the consistency of previous results in section \ref{pereversev_discussion}. \\
\\
Since $\hat{\mathcal{D}}$ is a second order differential operator, the complete set of solutions to $\hat{\mathcal{D}}$ can be decomposed into two separate families of solutions. The basic idea is to construct two separate families of operators, that we call $\hat{\mathcal{L}}$ and $\hat{\mathcal{R}}$\footnote{This stands for left and right, for lack of a better description of the operators. This is due to the fact that they will not generally function as exact inverses or adjoints of each other, which is demonstrated in appendix \ref{ladder_properties}.}, that commute with $\hat{\mathcal{D}}$. That is, $[\hat{\mathcal{D}}, \hat{\mathcal{L}}] = \hat{\mathcal{D}}\hat{\mathcal{L}} - \hat{\mathcal{L}}\hat{\mathcal{D}} = 0$, and similarly $[\hat{\mathcal{D}}, \hat{\mathcal{R}}] = 0$. Within this paragraph, $\hat{\mathcal{L}}$ and $\hat{\mathcal{R}}$ each refer to any operator from each family of operators, the elements of which will be elucidated in eq. \ref{eqn:Ladder_Operators}. These operators become useful if we know at least one particular solution $f$ to $\hat{\mathcal{D}}$. Using the commutativity with $\hat{\mathcal{L}}$ and $\hat{\mathcal{R}}$, this automatically implies that $\hat{\mathcal{L}}f$ and $\hat{\mathcal{R}}f$ are also solutions to $\hat{\mathcal{D}}$. We use this property to generate two families of solutions.  which are given by $\hat{\mathcal{L}}^nf$ and $\hat{\mathcal{R}}^nf$.\footnote{If $f$ is in the kernel of $\hat{\mathcal{L}}$ (or respectively $\hat{\mathcal{R}}$), we will only be able to generate the family corresponding to $\hat{\mathcal{R}}$ (respectively $\hat{\mathcal{L}}$). In that case, it will be necessary to find a second particular solution $f'$ to $\hat{\mathcal{D}}$, which will yield the second family of solutions.} For our current introductory purposes, we only present other important properties of our ladder operators in appendix \ref{ladder_properties}, where they are introduced in a more natural manner.\\
\\
Before constructing said ladder operators, it will be useful to consider the following corollary. Suppose that we define $\bm{\Psi}',\bm{\Psi},\bm{\Phi}$ to be complex-symmetric matrices obeying the following relation:
\begin{equation}
    \bm{\Psi}'=\bm{\Psi}+\bm{\Phi}.
\end{equation}
If we constrain both \(\bm{\Psi}'\) and \(\bm{\Psi}\) to obey eq. \ref{eqn:Beam_Tracing}, then the following must be true:
\begin{equation}
\begin{split}
&\frac{\mathrm{d}(\bm{\Psi}+\bm{\Phi})}{\mathrm{d}\tau}+(\bm{\Psi}+\bm{\Phi})\cdot\bm{\nabla}_{\bm{K}}\bm{\nabla}_{\bm{K}}H\cdot(\bm{\Psi}+\bm{\Phi})\\
&+\bm{\nabla}\bm{\nabla}_{\bm{K}}H\cdot(\bm{\Psi}+\bm{\Phi})+(\bm{\Psi}+\bm{\Phi})\cdot\bm{\nabla}_{\bm{K}}\bm{\nabla}H+\bm{\nabla}\bm{\nabla}H=0.
\end{split}
\end{equation}
Utilising eq. \ref{eqn:Beam_Tracing}, this time for \(\bm{\Psi}\), reduces this to:
\begin{equation}\label{eqn:Gaussian_Profile_Evolution}
\begin{split}
&\frac{\mathrm{d}\bm{\Phi}}{\mathrm{d}\tau}+\bm{\Phi}\cdot\bm{\nabla}_{\bm{K}}\bm{\nabla}_{\bm{K}}H\cdot\bm{\Phi}\\
&+\bm{\Psi}\cdot\bm{\nabla}_{\bm{K}}\bm{\nabla}_{\bm{K}}H\cdot\bm{\Phi}+\bm{\Phi}\cdot\bm{\nabla}_{\bm{K}}\bm{\nabla}_{\bm{K}}H\cdot\bm{\Psi}\\
&+\bm{\nabla}\bm{\nabla}_{\bm{K}}H\cdot\bm{\Phi}+\bm{\Phi}\cdot\bm{\nabla}_{\bm{K}}\bm{\nabla}H=0.
\end{split}
\end{equation}
This is equivalent to saying that if we have symmetric matrices $\bm{\Psi},\bm{\Phi}$ obeying eq. \ref{eqn:Beam_Tracing} and eq. \ref{eqn:Gaussian_Profile_Evolution} respectively, we know that their sum $\bm{\Psi}+\bm{\Phi}$ must also obey eq. \ref{eqn:Beam_Tracing}. This will turn out to be a very useful property in subsequent derivations.\\
\\
It is first instructive to analyse what happens if one inputs a Gaussian profile as our beam profile. This is not only an important consistency check for eq. \ref{eqn:Profile_Evolution}, but it also serves a crucial algebraic purpose later in our construction of the profiles generated by our ladder operators. 
\subsection{Gaussian profiles}
As we pointed out, one is free to decompose the profile in any way you like. Thus, even if our beam was exactly Gaussian, we could pull out a different Gaussian envelope, with the profile being the rest of our Gaussian beam(our profile is therefore Gaussian). We then expect our profile evolution equation to make our Gaussian profile evolve identically to the beam-tracing equations, which we subsequently prove. \\
\\
Let us suppose that we pick out a Gaussian envelope \(\bm{\Psi}(\tau)\), and our profile is given by:
\begin{equation}
    \mathcal{P}=\alpha(\tau)\exp\left(\frac{\mathrm{i}}{2}\bm{w}\cdot\bm{\Phi}(\tau)\cdot\bm{w}\right),
\end{equation}
Substituting the expression for our profile into eq. \ref{eqn:Profile_Evolution}, we will have terms proportional to \(\bm{ww}\), and terms that are independent. The terms independent of \(\bm{w}\) are:
\begin{equation}
    \frac{\mathrm{d}\ln(\alpha(\tau))}{\mathrm{d}\tau}+\frac{1}{2}\bm{\Phi } : \bm{\nabla}_{\bm{K}} \bm{\nabla}_{\bm{K}} H=0.
\end{equation}
The terms proportional to \(\bm{ww}\) are given by:
\begin{equation}
\begin{split}
\frac{\mathrm{i}}{2}\bm{ww}:\bigg(&\frac{\mathrm{d}\bm{\Phi}}{\mathrm{d}\tau}+\bm{\Phi}\cdot\bm{\nabla}_{\bm{K}}\bm{\nabla}_{\bm{K}}H\cdot\bm{\Phi}+(\bm{\Phi}\cdot\bm{\nabla}_{\bm{K}}\bm{\nabla}_{\bm{K}}H\cdot\bm{\Psi}+\bm{\Psi}\cdot\bm{\nabla}_{\bm{K}}\bm{\nabla}_{\bm{K}}H\cdot\bm{\Phi})\\
&+(\bm{\Phi}\cdot\bm{\nabla}_{\bm{K}}\bm{\nabla}H+\bm{\nabla}\bm{\nabla}_{\bm{K}}H\cdot\bm{\Phi})\bigg)=0.
\end{split}
\end{equation}
With this choice for $\mathcal{P}$, we see from eq. \ref{eqn:ansatz} and eq. \ref{eqn:amplitude_ansatz} that the total Gaussian envelope of the electric field, $\bm{\Psi}_\mathrm{tot}$, is given by the sum of the Gaussian envelope $\bm{\Psi}$, and the Gaussian envelope $\bm{\Phi}$ from $\mathcal{P}$.
\begin{equation}
    \bm{\Psi}_\mathrm{tot}=\bm{\Psi}+\bm{\Phi}.
\end{equation}
Notice that eq. \ref{eqn:Gaussian_Profile_Evolution} is precisely what our terms dependent on \(\bm{ww}\) tell us, albeit projected onto the \(\bm{ww}\)-plane. This confirms that if we were to have a Gaussian profile, the total envelope $\bm{\Psi}+\bm{\Phi}$ must obey eq. \ref{eqn:Beam_Tracing}. We can now turn our attention to the terms independent of \(\bm{w}\). Note that if we add eq. \ref{eqn:second_order_terms_no_w} to our terms, we end up with:
\begin{equation}\label{eqn:modified_amplitude_evolution}
    \frac{\mathrm{dln}(\alpha(\tau)A^{(0)}(\tau))}{\mathrm{d}\tau}+\frac{\mathrm{d}K_\mu}{\mathrm{d}\tau}\frac{\partial\hat{\bm{e}}}{\partial K_\mu}\cdot\hat{\bm{e}}^* +\hat{\bm{e}}^*\cdot\frac{\partial \bm{D}}{\partial K_\mu}\cdot\frac{\partial\hat{\bm{e}}}{\partial r_\mu}+\frac{1}{2}(\bm{\Psi}+\bm{\Phi}) : \bm{\nabla}_{\bm{K}} \bm{\nabla}_{\bm{K}} H=0.
\end{equation}
We will thus end up with the same amplitude evolution for \(\alpha(\tau)A^{(0)}\) instead of \(A^{(0)}\), using \(\bm{\Psi}+\bm{\Phi}\) instead of \(\bm{\Psi}\). Thus, the special case of applying a Gaussian beam profile to eq. \ref{eqn:Profile_Evolution} is consistent with our beam tracing equations.

\subsection{General solution -- ladder operators}
In this section, we will present the evolution of a generic starting profile $\mathcal{P}$. Our approach is similar to Leen et. al\cite{Leen2016}, but we instead construct generalised ladder operators for this system, which work with generic \(\tau\)-dependent tensor coefficients. This is a necessary feature in our case, as our tensor coefficients most definitely depend on \(\tau\). In a similar fashion to Leen et. al, in subsequent derivations we utilise the capitalised index `\textit{I}' to denote the ladder operator indexed by `\textit{I}'. Although this is certainly not necessary for the construction of the individual ladder operators, but it is useful book-keeping notation, and makes it explicitly clear that we expect to construct not just a single pair of ladder operators, but sets of ladder operator pairs. To be precise, in \textit{n}-dimensions, we require one set of \textit{n} independent ladder operators for $\hat{\mathcal{L}}_I$ which generate one family of solutions, and another such set to generate the other family of solutions from $\hat{\mathcal{R}}_I$. In the proceeding subsection, we will be outlining the definition and evolution of two ladder operators, \(\hat{\mathcal{L}}_I\) and  \(\hat{\mathcal{R}}_I\). It is clear from the context when `$\hat{\quad}$' refers to the ladder operator or a unit vector, as \(\hat{\mathcal{L}}_I\), \(\hat{\mathcal{R}}_I\) and \(\hat{\mathcal{D}}\) will be the main operators used subsequently. By the end of the section, we will have demonstrated that from appropriately chosen stationary states, we can generate families of solutions from either \(\hat{\mathcal{L}}_I\) or \(\hat{\mathcal{R}}_I\). In our particular case of \textit{n}=2, we will construct two ladder operators indexed by `\textit{I}', that is one pair for \(\hat{\mathcal{L}}_I\) and one pair for \(\hat{\mathcal{R}}_I\), where `\textit{I}' could be `\textit{x}' or `\textit{y}'. \\
\\
In this particular ladder operator approach, we are interested in ladder operators that commute with \(\hat{\mathcal{D}}\), unlike conventional ladder operator approaches. We will still obtain a ladder of states, but they will no longer be classified hierarchically by their eigenvalues, for example in Leen et. al\cite{Leen2016}, as our definition subsumes the exponential eigenvalue modulation into the solutions themselves.\\
\\
We first construct an operator of the form \(\hat{\mathcal{L}}_I=u_I^\mu(\tau)\frac{\partial}{\partial w_\mu}\), where \(u_I^\mu\) is an arbitrary vector. We want this operator to commute with \(\hat{\mathcal{D}}\), such that it can be used as a ladder operator to generate solutions of \(\hat{\mathcal{D}}\). The commutator of $\hat{\mathcal{L}}_I$ with \(\hat{\mathcal{D}}\) is given by: 
\begin{equation}
[\hat{\mathcal{D}},\hat{\mathcal{L}}_I]=\frac{\mathrm{d}u_I^\mu}{\mathrm{d}\tau}\frac{\partial}{\partial w_\mu}-u_I^\mu T_{w,\mu\nu}\frac{\partial}{\partial w_\nu}.
\end{equation}
In order for the commutator to evaluate to 0, our ladder operator components must satisfy the following ODE:
\begin{equation}
\frac{\mathrm{d}u_I^\nu}{\mathrm{d}\tau}=u_I^\mu T_{w,\mu\nu}.
\end{equation}
Our choice of \(\bm{u}_{I}(0)\) is still ultimately completely arbitrary, and depend on the specific boundary conditions we are interested in. After solving for \(\bm{u}_{I}(\tau)\), our ladder operator is given by:
\begin{equation}
    \hat{\mathcal{L}}_I(\tau)=\bm{u}_{I}(\tau)\cdot\bm{\nabla}_w.
\end{equation}
We can presciently construct another operator that will turn out to operate similar to an inverse operator. This operator turns out to be something of the form \(\hat{\mathcal{R}}_I=h_I^\nu(\tau)\frac{\partial}{\partial w_\nu}+v^\nu_I(\tau)w_\nu\), where both \(v_I^\nu\) and \(h_I^\nu\) are arbitrary vectors. We find that:
\begin{equation}
    [\hat{\mathcal{D}},\hat{\mathcal{R}}_I]=\frac{\mathrm{d}v_I^\mu}{\mathrm{d}\tau}w_\mu+\frac{\mathrm{d}h_I^\mu}{\mathrm{d}\tau}\frac{\partial}{\partial w_\mu}-\mathrm{i}\frac{\partial}{\partial K_\mu} \frac{\partial}{\partial K_\nu} Hv_I^\mu\frac{\partial}{\partial w_\nu}+w_\mu T_{\mu\nu}v_I^\nu-h_I^\mu T_{\mu\nu}\frac{\partial}{\partial w_\nu}.
\end{equation}
Comparing the coefficients of $\frac{\partial}{\partial w_\nu}$ and $w_\nu$, we require each set of coefficients to be 0. We thus find that \(v^\nu_I(\tau)\) must be given by:
\begin{equation}\label{eqn:vector_evolution}
    \frac{\mathrm{d}v_I^\mu}{\mathrm{d}\tau}=-T_{w,\mu\nu}v_I^\nu.
\end{equation}
We also get an ODE for the evolution of \(h^\nu_I(\tau)\):
\begin{equation}\label{eqn:hI_evolution}
    \frac{\mathrm{d}h_I^\mu}{\mathrm{d}\tau}=h_I^\nu T_{w,\nu\mu}+\mathrm{i}v_I^\nu\frac{\partial}{\partial K_\nu} \frac{\partial}{\partial K_\mu} H.
\end{equation}
The subsequent mathematics will be much more illuminating if we take \(h_I^\mu(\tau)=M_{\mu\nu}(\tau)v^\nu_I(\tau)\), as this will expose a deep link to the Gaussian profile. Using eqs. \ref{eqn:vector_evolution} \& \ref{eqn:hI_evolution}, we then arrive at another ODE for \(M_{\mu\nu}\):
\begin{equation}
    \left(\frac{\mathrm{d}M_{\mu\nu}}{\mathrm{d}\tau}\right)_w=\left(\mathrm{i}\frac{\partial}{\partial K_\nu} \frac{\partial}{\partial K_\mu} H+M_{\rho\nu}T_{w,\rho\mu}+M_{\mu\rho}T_{w,\rho\nu}\right)_w
\end{equation}
We can choose $\bm{M}$ to be a symmetric matrix. Absorbing the the terms involving $\left({\mathrm{d}\hat{\bm{y}}}/{\mathrm{d}\tau}\cdot\hat{\bm{x}}\right)(\hat{\bm{x}}\hat{\bm{y}}-\hat{\bm{y}}\hat{\bm{x}})$ in our expression for $\bm{T}_w$ into our definition for ${\mathrm{d}\bm{M}}/{\mathrm{d}\tau}$ in tensor notation, we find the following simplified expression, where our total derivative includes the changes in the tensor due to the \(\tau\)-dependent basis $(\hat{\bm{x}},\hat{\bm{y}})$:
\begin{equation}\label{eqn:ODE_for_M}
\begin{split}
\left(\frac{\mathrm{d}\bm{M}}{\mathrm{d}\tau}\right)_w&=\big(\mathrm{i}\bm{\nabla}_{\bm{K}}\bm{\nabla}_{\bm{K}}H\\
&+\bm{M}\cdot\bm{\Psi}\cdot\bm{\nabla}_{\bm{K}}\bm{\nabla}_{\bm{K}}H+\bm{\nabla}_{\bm{K}}\bm{\nabla}_{\bm{K}}H\cdot\bm{\Psi}\cdot\bm{M}\\
&+\bm{M}\cdot\bm{\nabla}\bm{\nabla}_{\bm{K}}H+\bm{\nabla}_{\bm{K}}\bm{\nabla}H\cdot\bm{M}\big)_w,
\end{split}
\end{equation}
Without loss of generality, if \(\bm{M}\) obeys the unprojected equation, this would imply that it is a solution to eq. \ref{eqn:ODE_for_M} as well. Specifically, if \(\bm{M}\) satisfies:
\begin{equation}
\begin{split}
\frac{\mathrm{d}\bm{M}}{\mathrm{d}\tau}&=\mathrm{i}\bm{\nabla}_{\bm{K}}\bm{\nabla}_{\bm{K}}H\\
&+\bm{M}\cdot\bm{\Psi}\cdot\bm{\nabla}_{\bm{K}}\bm{\nabla}_{\bm{K}}H+\bm{\nabla}_{\bm{K}}\bm{\nabla}_{\bm{K}}H\cdot\bm{\Psi}\cdot\bm{M}\\
&+\bm{M}\cdot\bm{\nabla}\bm{\nabla}_{\bm{K}}H+\bm{\nabla}_{\bm{K}}\bm{\nabla}H\cdot\bm{M},
\end{split}
\end{equation}
then it is a solution to eq. \ref{eqn:ODE_for_M}.
Note that this bears a strong resemblance to our equation for the evolution of a Gaussian profile, eq. \ref{eqn:Gaussian_Profile_Evolution}. In fact it is almost identical to the equation for the evolution of the inverse profile, given by:
\begin{equation}
\begin{split}
\frac{\mathrm{d}\bm{\Phi}^{-1}}{\mathrm{d}\tau}&=\bm{\nabla}_{\bm{K}}\bm{\nabla}_{\bm{K}}H\\
&+\bm{\Phi}^{-1}\cdot\bm{\Psi}\cdot\bm{\nabla}_{\bm{K}}\bm{\nabla}_{\bm{K}}H+\bm{\nabla}_{\bm{K}}\bm{\nabla}_{\bm{K}}H\cdot\bm{\Psi}\cdot\bm{\Phi}^{-1}\\
&+\bm{\Phi}^{-1}\cdot\bm{\nabla}\bm{\nabla}_{\bm{K}}H+\bm{\nabla}_{\bm{K}}\bm{\nabla}H\cdot\bm{\Phi}^{-1}.
\end{split}
\end{equation}
thus, \(\bm{M}=\mathrm{i}\bm{\Phi}^{-1}\) is a valid solution in general, for any \(\bm{\Phi}\) where both \(\bm{\Psi}+\bm{\Phi}\) and \(\bm{\Psi}\) obey the beam tracing equations. Note that \(\bm{\Phi}\) need not be full-rank, due to the fact that we only need to satisfy the ODE for the components along the width of the beam. We get no further information from the components that do not lie along the width of the beam at any rate. In the case that \(\bm{\Phi}\) is not full-rank, \(\bm{\Phi}^{-1}\) should be interpreted as the Moore-Penrose inverse. It would not be full rank if we use the prescription in eq. \ref{eqn:smart_envelope_choice}, which would be cleanest prescription to use from a computational standpoint.\\
\\
To summarise, our ladder operators are: 
\begin{equation}\label{eqn:Ladder_Operators}
\begin{split}
    \hat{\mathcal{L}}_I&=\bm{u}_{I}(\tau)\cdot\bm{\nabla}_w,\\
    \hat{\mathcal{R}}_I&=\bm{v}_{I}(\tau)\cdot(\bm{w}+\mathrm{i}\bm{\Phi}^{-1}\cdot\bm{\nabla}_w).
\end{split}
\end{equation}
These two families of operators commute with our differential operator, allowing us to exactly calculate the evolution of any beam profile matching their boundary conditions. For example, in this case our solutions of eq. \ref{eqn:Profile_Evolution} will have to be two-dimensional in a chosen basis $(\hat{\bm{x}},\hat{\bm{y}})$. Thus, we minimally require two operators $\hat{\mathcal{L}}_x$ and $\hat{\mathcal{L}}_y$ to generate one family of solutions, and another pair of operators $\hat{\mathcal{R}}_x$ and $\hat{\mathcal{R}}_y$ to generate the other family of solutions.\\
\\
As a final note, we want to point out that we are free to choose any convenient symmetric \(\bm{\Psi}_\mathrm{tot}\) for obtaining \(\bm{\Phi}\), as the only thing we have set in stone is \(\bm{\Psi}\) when we defined our beam envelope. The initial choice of \(\bm{\Phi}\) simply affects the initial definition for \(\bm{\Psi}_\mathrm{tot}\) and vice versa. In particular, note that if we make the following definition:
\begin{equation}\label{eqn:smart_envelope_choice}
\begin{split}
    \bm{\Psi}&=\bm{\Psi}_w+\frac{\hat{\bm{g}}}{g}\frac{\mathrm{d}\bm{K}}{\mathrm{d}\tau}+\left(\frac{\mathrm{d}\bm{K}}{\mathrm{d}\tau}\right)_w\frac{\hat{\bm{g}}}{g},\\
    \bm{\Psi}_\mathrm{tot}&=\bm{\Psi}_{w,\mathrm{tot}}+\frac{\hat{\bm{g}}}{g}\frac{\mathrm{d}\bm{K}}{\mathrm{d}\tau}+\left(\frac{\mathrm{d}\bm{K}}{\mathrm{d}\tau}\right)_w\frac{\hat{\bm{g}}}{g},
\end{split}
\end{equation}
then we are assured that \(\bm{\Phi}=\bm{\Psi}_{w,\mathrm{tot}}-\bm{\Psi}_w\). This is a neat choice as we do not care about anything other than the components along \(\bm{ww}\) in the first place, and will henceforth be the choice of \(\boldsymbol{\Phi}\) we use in the rest of the manuscript. \\
\\
Firstly, we have to identify the stationary states of our ladder operators, so we look for the two functions for which \(\hat{\mathcal{L}}_If=0\) and \(\hat{\mathcal{R}}_Ig=0\). On top of this, the solutions need to be solutions to $\mathcal{D}$.\footnote{Notice that if we had a function \textit{f} satisfying $\hat{\mathcal{L}}_If=0$, then it is clear that $\hat{\mathcal{D}}\hat{\mathcal{L}}_If=\hat{\mathcal{L}}_I\hat{\mathcal{D}}f=0$. It is easy to see that $\hat{\mathcal{L}}_I$ and $\hat{\mathcal{R}}_I$ only have one-dimensional kernels, thus this tells us that $\hat{\mathcal{D}}f=C(\tau)f$. Hence, stationary states of our ladder operators are almost solutions to $\hat{\mathcal{D}}$ in general, and can be made into solutions by redefining to $f'=f\exp\left(\int_0^\tau C(\tau')\mathrm{d}\tau'\right)$, such that $\hat{\mathcal{D}}f'=0$. The same argument trivially follows for $\hat{\mathcal{R}}_I$.} These are \(f=C\) for \(\hat{\mathcal{L}}_I\) and \(g=\alpha(\tau)\exp\left(({\mathrm{i}}/{2})\bm{w}\cdot\bm{\Phi}\cdot\bm{w}\right)\) for \(\hat{\mathcal{R}}_I\), which are precisely the constant profile and Gaussian profile solutions we were already aware of, suggesting that we can use them to generate a family of solutions to $\hat{\mathcal{D}}$ each. We generate each family of solutions by either repeated application of various \(\hat{\mathcal{R}}_I\) to the constant profile solution, or repeated application of various \(\hat{\mathcal{L}}_I\) to the Gaussian profile solution. In particular, we can generate each family of solutions, $f_{nm}$ and $g_{nm}$, in our two-dimensional case(see section \ref{GH_Boundary} for details) using:
\begin{equation}
\begin{split}
    f_{nm}&=\hat{\mathcal{R}}^n_x\hat{\mathcal{R}}^m_y(1),\\
    g_{nm}&=\alpha(\tau)\hat{\mathcal{L}}^n_x\hat{\mathcal{L}}^m_y\left(\exp\left(\frac{\mathrm{i}}{2}\bm{w}\cdot\bm{\Phi}\cdot\bm{w}\right)\right),\\
\end{split}
\end{equation}
where we choose \(\alpha(0)=1\).\\
\\
To summarise, in this section we have found ladder operator solutions to $\hat{\mathcal{D}}$. In the subsequent subsection, we will utilise the ladder operators to generate solutions which reduce to the Hermite polynomials at the plasma boundary, $\tau=0$. The Hermite polynomial profile is a useful boundary profile as it allows us to easily decompose a complicated electric field into various Gauss-Hermite modes by utilising the fact that the Gauss-Hermite modes form an orthogonal basis. In theory, this allows us to model the evolution of an arbitrary profile by first decomposing it into its Gauss-Hermite mode constituents, and then propagating each mode.
\subsection{Gauss-Hermite boundary conditions\label{GH_Boundary}}
As mentioned earlier, we would like these profiles to reduce to Hermite polynomials at $\tau=0$. We can do this simply by making appropriate choices for \(\bm{u}_{I}(0)\), \(\bm{v}_{I}(0)\), \(\bm{\Phi}\) and \(\bm{\Psi}\). We are free to choose any arbitrary scaling for our Hermite polynomials due to this freedom, which we will utilise to choose a neat set of boundary conditions in section \ref{trad_hermite}, in this case yielding physicist's Hermite polynomials instead of probabilist's Hermite polynomials. However, the subsequent prescription we provide reduces to the `elegant' Hermite polynomials\cite{pampaloni2004} at the plasma boundary, which are probabilist's Hermite polynomials. 
This will serve as a useful demonstration of our approach, and it serves as a natural basis for decomposing an arbitrary electric field at the boundary. To do this, we will first express the elegant Hermite polynomials(or any Hermite polynomial convention of one's choosing) in terms of ladder operators, then match the ladder operators we developed.

\subsubsection{Case 1}
Suppose that we want to generate probabilist's Hermite polynomials of a dimensionless variable $\bm{x}$ at the boundary by the following ladder operators, where \(\hat{\bm{e}}_i\) is a unit vector in a given coordinate basis and $\boldsymbol{\nabla}_x$ is the gradient operator in $\bm{x}$:
\begin{equation}
\begin{split}
    \hat{A}_i^-&=\hat{\bm{e}}_i\cdot\bm{\nabla}_x,\\
    \hat{A}_i^+&=\hat{\bm{e}}_i\cdot(\bm{x}-\bm{\nabla}_x).\\
\end{split}
\end{equation}
The \textit{n}-th order probabilist's Hermite polynomial in the coordinate \textit{x} is given by \(H_n(x)=\hat{A}_x^{+,n}(1)\), referring to application of \(\hat{A}_i^{+}\) \textit{n} times on unity. In 2D, with two independent coordinates \textit{x} and \textit{y}, the $(n,m)$-th order Hermite polynomial is unambiguously given by \(H_n(x)H_m(y)=\hat{A}_x^{+,n}\hat{A}_y^{+,m}(1)\), where we can apply the ladder operators in any order due to the fact that the ladder operators in orthogonal directions commute(see section \ref{ladder_properties} for details). \\
\\
We first convert the ladder operators \(\hat{\mathcal{L}}_I\) and \(\hat{\mathcal{R}}_I\) to dimensionless coordinates \(\bm{x}(\tau)=\sqrt{\mathrm{i}\bm{\Phi}}\cdot\bm{w}\), in order to directly compare them with the aforementioned ladder operators, \(\hat{A}_x^{-}\) and \(\hat{A}_x^{+}\) respectively. We then have:
\begin{equation}
\begin{split}
    \hat{\mathcal{L}}_I&=\bm{u}_{I}(\tau)\cdot\sqrt{\mathrm{i}\bm{\Phi}}\cdot\bm{\nabla}_x,\\
    \hat{\mathcal{R}}_I&=\bm{v}_{I}(\tau)\cdot\sqrt{\mathrm{i}\bm{\Phi}}^{-1}\cdot(\bm{x}-\bm{\nabla}_x),\\
\end{split}
\end{equation}
where \(\sqrt{\bm{\Phi}}\) is defined to be \(\sqrt{\bm{\Phi}}\cdot\sqrt{\bm{\Phi}}=\bm{\Phi}\). This is not always possible as \(\bm{\Phi}\) is in general a 2D complex-symmetric tensor, but given that we are only trying to match to boundary conditions, we take \(\bm{\Phi}(0)\) to be a diagonalisable symmetric tensor. We could for example, choose it to be $\Re(\bm{\Psi}_{w,0})$ or $\Im(\bm{\Psi}_{w,0})$, but in general it can be any symmetric tensor appropriately chosen to recover the desired boundary conditions. In order to obtain the elegant Hermite polynomial basis for our Gauss-Hermite beams, we set \(\bm{x}(0)=\sqrt{-\mathrm{i}\bm{\Psi}_{w,0}}\cdot\bm{w}\). We implicitly assume that \(\bm{\Psi}_{w,0}\) is diagonalisable such that we are assured that the square root exists. Without loss of generality, we will define our Hermite polynomial boundary profile in the basis $(\hat{\bm{
x}},\hat{\bm{
y}})$ that diagonalises \(\bm{\Psi}_{w,0}\)(such a basis must be orthonormal for any diagonalisable symmetric matrix), such that we construct a basis of Gauss-Hermite modes that allows one to decompose a given boundary electric field into Gauss-Hermite modes. We can then easily choose boundary conditions to recover the probabilist's Hermite polynomials in $\bm{x}(0)$ at the boundary.\\
\\
We set \(\bm{\Phi}(\tau=0)=-\bm{\Psi}_{w,0}\) in order to reproduce the ladder operators for the Gauss-Hermite profiles at the boundary. This means that \(\bm{\Psi}_\mathrm{tot}(\tau=0)=0\), so we thus just need to evolve this alongside \(\bm{\Psi}\) to be able to keep track of the evolution of \(\bm{\Phi}\). Given the diagonalised basis of \(\bm{\Psi}_{w,0}\), where \(\{\hat{\bm{e}}_I\}=(\hat{\bm{
x}},\hat{\bm{
y}})\), and $\Psi_{II,0}$(`\textit{I}' refers to \textit{x} or \textit{y}) refers to the components of \(\bm{\Psi}_{w,0}\) in the diagonal basis, we then define:
\begin{equation}
\begin{split}
    \bm{u}_I(0)&=\sqrt{\mathrm{i}\Psi_{II,0}}^{-1}\hat{\bm{e}}_I,\\
    \bm{v}_I(0)&=\sqrt{\mathrm{i}\Psi_{II,0}}\hat{\bm{e}}_I,\\
\end{split}
\end{equation}
in order for our ladder operators to reduce to that of the normalised elegant Hermite polynomials. With these choices, the solution given by:
\begin{equation}
    f_{nm}=\hat{\mathcal{R}}_x^n\hat{\mathcal{R}}_y^m(1),
\end{equation}
satisfies the boundary condition \(f_{nm}(\tau=0)=H_n\left(\sqrt{-\mathrm{i}\Psi_{0,xx}}w_x\right)H_m\left(\sqrt{-\mathrm{i}\Psi_{0,yy}}w_y\right)\).
\subsubsection{Case 2}
We could instead generate our Hermite polynomials at the boundary by the following ladder operators:
\begin{equation}\label{eqn:ladder_ops_case2}
\begin{split}
    \hat{B}_i^+&=-\hat{\bm{e}}_i\cdot\bm{\nabla}_x,\\
    \hat{B}_i^-&=\hat{\bm{e}}_i\cdot(\bm{x}+\bm{\nabla}_x).\\
\end{split}
\end{equation}
where in a similar fashion to the preceding case, \(h_n(x)h_m(y)=\hat{B}_x^{+,n}\hat{B}_y^{+,m}\exp\left(-\frac{1}{2}\bm{x}\cdot\bm{x}\right)\). As follows from our earlier definition, the Hermite functions are simply related to the Hermite polynomials by the following:
\begin{equation}
    h_n(x)h_m(y)\exp\left(\frac{1}{2}\bm{x}\cdot\bm{x}\right)=H_n(x)H_m(y)=\exp\left(\frac{1}{2}\bm{x}\cdot\bm{x}\right)\hat{L}_x^{+,n}\hat{L}_y^{+,m}\exp\left(-\frac{1}{2}\bm{x}\cdot\bm{x}\right).
\end{equation}
Similarly to the previous case, we convert to dimensionless coordinates \(\bm{x}(\tau)=\sqrt{-\mathrm{i}\bm{\Phi}}\cdot\bm{w}\), and we once again define \(\bm{x}(0)=\sqrt{-\mathrm{i}\bm{\Psi}_{w,0}}\cdot\bm{w}\). To obtain Hermite functions at the boundary, we once again have to match $\hat{\mathcal{L}}_I$ and $\hat{\mathcal{R}}_I$ to our previously defined ladder operators in eq. \ref{eqn:ladder_ops_case2}:
\begin{equation}
\begin{split}
    \hat{\mathcal{L}}_I&=\bm{u}_{I}(\tau)\cdot\sqrt{-\mathrm{i}\bm{\Phi}}\cdot\bm{\nabla}_x,\\
    \hat{\mathcal{R}}_I&=\bm{v}_{I}(\tau)\cdot\sqrt{-\mathrm{i}\bm{\Phi}}^{-1}\cdot(\bm{x}+\bm{\nabla}_x),\\
\end{split}
\end{equation}
We simply need to set \(\bm{\Phi}(\tau=0)=\bm{\Psi}_{w,0}\) and \(\bm{\Psi}(\tau=0)=0\), such that \(\bm{\Psi}_\mathrm{tot}(\tau=0)=\bm{\Psi}_{w,0}\). Once again, we have to assume that $\bm{\Psi}_{w,0}$ is diagonalisable and that we work in the diagonalised basis of \(\bm{\Psi}_{w,0}\). $\Psi_{II,0}$ and \(\hat{\bm{e}}_I\) are defined identically to case 1, so we then set:
\begin{equation}
\begin{split}
    \bm{u}_I(0)&=-\sqrt{\mathrm{i}\Psi_{II,0}}^{-1}\hat{\bm{e}}_I,\\
    \bm{v}_I(0)&=\sqrt{\mathrm{i}\Psi_{II,0}}\hat{\bm{e}}_I,\\
\end{split}
\end{equation}
in order for our ladder operators to reduce to that of the normalised elegant Hermite functions. Thus, if $\alpha(\tau)$ is the same Gaussian profile amplitude function that we solved for in eq. \ref{eqn:modified_amplitude_evolution}, the solution given by:
\begin{equation}
    g_{nm}=\alpha(\tau)\hat{\mathcal{L}}_x^n\hat{\mathcal{L}}_y^m\exp\left(\frac{\mathrm{i}}{2}\bm{w}\cdot\bm{\Phi}\cdot\bm{w}\right),
\end{equation}
reduces to:
\begin{equation}
    g_{nm}(\tau=0)=H_n\left(\sqrt{-\mathrm{i}\Psi_{0,xx}}w_x\right)H_m\left(\sqrt{-\mathrm{i}\Psi_{0,yy}}w_y\right)\alpha(0)\exp\left(\frac{\mathrm{i}}{2}\bm{w}\cdot\bm{\Psi}_{w,0}\cdot\bm{w}\right).
\end{equation}
Tying everything together, we now show that both choices can be chosen to be consistent with each other and reduce to the same boundary conditions. Note that since \(\exp\left(({\mathrm{i}}/{2})\bm{w}\cdot\bm{\Phi}\cdot\bm{w}\right)\exp\left(({\mathrm{i}}/{2})\bm{w}\cdot\bm{\Psi}\cdot\bm{w}\right)=\exp\left(({\mathrm{i}}/{2})\bm{w}\cdot\bm{\Psi}_\mathrm{tot}\cdot\bm{w}\right)\), but in both cases, \(\bm{\Psi}_\mathrm{tot}(\tau=0)=\bm{\Psi}_0\), so the solutions given by either approach will be identical. With both cases, the electric field given by the total beam, \(\mathcal{P}A^{(0)}\exp(\mathrm{i}\psi)\), match the exact same boundary conditions. If both cases have the same boundary conditions, then by uniqueness of solutions to eq. \ref{eqn:Profile_Evolution}, we are assured that they must be the same throughout the plasma. Thus, we have outlined two cases which have been defined to reduce to the elegant Gauss-Hermite modes at $\tau=0$. \\
\\
Importantly, note that the opposite definitions of \(\bm{\Psi}_\mathrm{tot}\) and \(\bm{\Psi}\) in each case, along with the fact that \(\bm{\Phi}(\tau)=\bm{\Psi}_\mathrm{tot}-\bm{\Psi}\), means that \(\bm{\Phi}(\tau)^\mathrm{(Case\,1)}=-\bm{\Phi}(\tau)^\mathrm{(Case\,2)}\). This crucial distinction guarantees that our two approaches yield identical results in both cases. Therefore, which approach one chooses ultimately depends on personal preference.\\
\\
At this juncture, we will further expound on the `ladder' operation performed by the ladder operators. As we will show in appendix \ref{ladder_properties}, the ladder operators as we have just defined actually function as inverses of each other. If we had a solution $f_{nm}=\hat{\mathcal{R}}_x^n\hat{\mathcal{R}}_y^m(1)$, then $\hat{\mathcal{L}}_x^i\hat{\mathcal{L}}_y^jf_{nm}\propto\hat{\mathcal{R}}_x^{n-i}\hat{\mathcal{R}}_y^{m-j}(1)=f_{(n-i)(m-j)}$. Thus, we have a ladder of solutions which our operators are capable of lowering and raising us through, which is where the terminology arises from.\\
\\
In order to analyse the solution at generic $\tau$, not just at $\tau=0$, we have to treat the square root of \(\bm{\Phi}\) with some nuance. In general, one can take the square-root of any diagonalisable tensor with no issues, save for an ambiguity in sign. It is easy to see that this sign ambiguity will have no effect on our results. However, the square-root of a generic tensor is not defined, thus it may not always be possible to take the square root. However, one should not lose sight of the fact that \(\sqrt{\bm{\Phi}}\) is only ever used as a coordinate transformation. The actual expression for our ladder operators do not involve square roots, hence if there is a case in which \(\bm{\Phi}\) cannot be square-rooted, direct calculation from the ladder operators is still perfectly feasible and well-defined. In the specific case of Gauss-Hermite beam tracing, we want sensible boundary conditions for incident Gauss-Hermite beams, for which we only require square root-able \(\bm{\Phi}(0)\), which we are ultimately free to specify. For example, we are assured that the real and imaginary parts of a generic $\bm{\Psi}_{w,0}$ are separately diagonalisable, and hence we can take the square roots of those, even if $\bm{\Psi}_{w,0}$ itself is not diagonalisable. Using that fact, one can always define traditional Hermite polynomial solutions at the boundary if all else fails. For example, one can always set \(\bm{\Phi}(0)=\pm2\mathrm{i}\Im(\bm{\Psi}_w)(0)\), depending on which ladder operator we are using to generate the solutions(`+' for case 2 and `-' for case 1), and adjust the initial condition for \(\bm{\Psi}_w\) accordingly. We are always able to decompose any boundary electric field into components in the Gauss-Hermite beams with \(\bm{x}=\sqrt{2\Im(\bm{\Psi}_w)(0)}\cdot\bm{w}\) at the boundary. Section \ref{trad_hermite} goes into more detail for applying this in vacuum, where this specific choice has the added advantage of being diagonal in the energy basis. That is, the traditional Hermite modes importantly allow us to decompose the energy into contributions from each Hermite mode, which one is not able to do in the `elegant' Hermite polynomial basis.
\subsection{Hermite tensors}\label{hermite_tensors}
Assuming that said square root can indeed be taken, we would like to point out the link between our solutions and the Hermite tensors\cite{grad1949,maheshwari2014}. We have essentially derived the coefficient tensor with which we contract our Hermite tensors, where the Hermite tensors are given by:
\begin{equation}\label{eqn:Hermite_tensor_ladder_ops}
    \bm{H}^{(N)}=(\bm{x}-\bm{\nabla}_x)^N (1).
\end{equation}
A similar expression can be given using the Gaussian generating function, which we do not present here as it has no use to us. Before discussing how this relates to our solution, we first want to give a short argument that prescribes the exact expression for Hermite tensors of order \textit{N}. Firstly, from our ladder operator expression, it is clear that the \textit{N}-index Hermite tensor must be symmetric in its \textit{N} indices. Furthermore, if we choose diagonal components of our Hermite tensor, they must reduce to the \textit{N}-th order Hermite polynomial of the coordinate of that diagonal. That is: 
\begin{equation}
    H^{(N)}_{\mu\mu...\mu}=H_N(x_\mu).
\end{equation}
By combining these properties, as well as noting that eq. \ref{eqn:Hermite_tensor_ladder_ops} suggests that our \textit{N}-th order Hermite tensor is a tensor product of \(x_\mu\) and \(\delta_{\mu\nu}\), we can conclude that there is only one possible procedure that results in this. Consider the \textit{N}-th order Hermite polynomial, it will have an expression of the form:
\begin{equation}
    H_N(x)=a_0x^N+a_1x^{N-2}+...a_kx^{N-2k}+...+a_{\left(\frac{N-N\mathrm{mod(2)}}{2}\right)}x^{N\mathrm{mod(2)}}.
\end{equation}
In order for the diagonal elements to reduce to this whilst being symmetric in all indices, this is the only possible form for the Hermite tensor:
\begin{equation}
    H^{(N)}_{\mu_1...\mu_N}(x)=\sum^{\left(\frac{N-N\mathrm{mod(2)}}{2}\right)}_{k=0}a_k\underbrace{x_{\{\mu_1}...x_{\mu_{N-2k}}}_{N-2k\,\mathrm{terms}}\underbrace{\delta_{\mu_{N-2k-1}\mu_{N-2k-2}}...\delta_{\mu_{N-1}\mu_{N}\}}}_{k\,\mathrm{terms}},
\end{equation}
where `\(_{\{\mu_1\mu_2...\mu_N\}}\)' refers to the symmetrisation operation over the indices. In order to obtain explicit polynomials solutions, we have to contract this expression with some coefficient tensor. The coefficient tensor we have then found is just:
\begin{equation}
    \bm{C}^{(N)}=\prod_{i=0}^N \left(\bm{v}_{I}(\tau)\cdot\sqrt{\mathrm{i}\bm{\Phi}}^{-1}\right),
\end{equation}
where the products here are strictly tensor products. Our solutions are simply:
\begin{equation}\label{eqn:HermiteTensorSolution}
    f_{{n_0}{n_1}{n_2}...{n_N}}=\bm{C}^{(N)} \overset{\text{full}}{:} \bm{H}^{(N)}.
\end{equation}
Note that in the final expressions, we will not encounter any square-roots, so the expression we obtain is well-defined, as one would expect as our ladder operators did not involve square-roots, and they are only used as mathematical machinery for algebraic manipulation here. This relationship with the Hermite tensors is indeed incredibly useful for our purposes, as it would speed up application of ladder operators, as we have demonstrated that it amounts to tensor contraction with the Hermite tensor, thus all one needs to simulate is the evolution of $\bm{v}_I$. 

\subsection{Exact Hermite polynomial solutions}
With all these properties of our ladder operators established, we will now investigate the generic evolution of a Hermite polynomial initial condition. Since both cases will yield the same result, we will focus on the approach provided by case 1. Assuming that \(\bm{x}=\sqrt{\mathrm{i}\bm{\Phi}}\cdot\bm{w}\) exists, we see that our ladder operators take the form:
\begin{equation}
\begin{split}
    \hat{\mathcal{L}}_I&=\bm{u}_{I}(\tau)\cdot\bm{\nabla}_w,\\
    &= \bm{\mu}_{I}(\tau)\cdot\bm{\nabla}_x,\\
    \hat{\mathcal{R}}_I&=\bm{v}_{I}(\tau)\cdot(\bm{w}+\mathrm{i}\bm{\Phi}^{-1}\cdot\bm{\nabla}_w),\\
    &=\bm{\nu}_{I}(\tau)\cdot(\bm{x}-\bm{\nabla}_x),
\end{split}
\end{equation}
where $\bm{\mu}_{I}=\bm{u}_{I}(\tau)\cdot\sqrt{\mathrm{i}\bm{\Phi}}$ and $\bm{\nu}_{I}=\bm{v}_{I}(\tau)\cdot\sqrt{\mathrm{i}\bm{\Phi}}^{-1}$
In general, these need not be orthogonal sets of vectors. The profile which reduces to the 2D \((n,m)\) Hermite polynomial mode is given by:
\begin{equation}
    f_{nm}(\bm{x})=\hat{\mathcal{R}}_x^n \hat{\mathcal{R}}_y^m(1).
\end{equation}
If \(\bm{v}_{I}(\tau)\) are not orthogonal, then we would not have a single Hermite mode as the solution. Given that we only have a 2D problem, this is still quite tractable. Recall that \(\bm{\nu}_I=\bm{v}_I\cdot\sqrt{\mathrm{i}\bm{\Phi}}^{-1}\). Suppose that we have some generic \(\bm{\nu}_x\) and \(\bm{\nu}_y\). Choosing \(\hat{\bm{e}}_1=\hat{\bm{\nu}}_x\) to be one of the principal axes for our Hermite polynomials, we then define \(\hat{\bm{e}}_2=\frac{(1-\hat{\bm{\nu}}_x\hat{\bm{\nu}}_x)\cdot\hat{\bm{\nu}}_y}{\sqrt{1-(\hat{\bm{\nu}}_x\cdot\hat{\bm{\nu}}_y)^2}}\), which is the normalised projection of \(\hat{\bm{\nu}}_y\) perpendicular to \(\hat{\bm{\nu}}_x\). Our ladder operators thus become:
\begin{equation}
\begin{split}
    \hat{\mathcal{R}}_x&=|\bm{\nu}_x|\hat{\mathcal{R}}_1,\\
    \hat{\mathcal{R}}_y&=\bm{\nu}_y\cdot\hat{\bm{e}}_1\hat{\mathcal{R}}_1+\bm{\nu}_y\cdot\hat{\bm{e}}_2\hat{\mathcal{R}}_2
\end{split}
\end{equation}
where:
\begin{equation}
\begin{split}
    \hat{\mathcal{R}}_1&=\hat{\bm{e}}_1\cdot(\bm{x}-\bm{\nabla}_x),\\
    \hat{\mathcal{R}}_2&=\hat{\bm{e}}_2\cdot(\bm{x}-\bm{\nabla}_x).\\
\end{split}
\end{equation}
\(\hat{\mathcal{R}}_1\) and \(\hat{\mathcal{L}}_2\) generate the Hermite polynomials in the basis given by \(\hat{\bm{e}}_1\) and \(\hat{\bm{e}}_2\). Therefore:
\begin{equation}
    f_{nm}(\bm{x})=|\bm{\nu}_x|^n\hat{\mathcal{R}}^n_1(\bm{\nu}_y\cdot\hat{\bm{e}}_1\hat{\mathcal{R}}_1+\bm{\nu}_y\cdot\hat{\bm{e}}_2\hat{\mathcal{R}}_2)^m(1).
\end{equation}
Expressing this explicitly in terms of Hermite polynomials, we have:
\begin{equation}\label{eqn:profile_exact_solution}
    f_{nm}(\bm{x})=|\bm{\nu}_x|^n\sum_{k=0}^m\left(\left(\begin{array}{cc} m\\ k \end{array}\right)(\bm{\nu}_y\cdot\hat{\bm{e}}_1)^{m-k}(\bm{\nu}_y\cdot\hat{\bm{e}}_2)^k H_{n+m-k}(\hat{\bm{e}}_1\cdot\sqrt{\mathrm{i}\bm{\Phi}}\cdot\bm{w})H_{k}(\hat{\bm{e}}_2\cdot\sqrt{\mathrm{i}\bm{\Phi}}\cdot\bm{w})\right)
\end{equation}
This quantity is well defined after we expand out everything, as we will not have any square roots of matrices left, in a similar fashion to what happens in eq. \ref{eqn:HermiteTensorSolution}. We thus see that in general, an \(nm\)-Hermite mode will evolve into a linear combination of Hermite modes in a rotated basis. Note that a simple product of Hermite polynomials can always be expressed as a linear combination of Hermite polynomials in some other rotated basis, hence we cannot presently conclude that the solution can only be expressed as a linear combination of Hermite modes. In homogeneous media this is clear cut, as \(\bm{\nu}_x\) and \(\bm{\nu}_y\) remain in the same directions throughout the entire evolution, thus, an \(nm\)-Hermite mode remains a single \(nm\)-Hermite mode.\\
\\
At the end of the day, one could always apply the ladder operators directly to evaluate the resulting polynomial, avoiding the need to deal with a sum of products of Hermite polynomials, which somewhat defeats the purpose of having Hermite polynomials. However, note that although a linear combination of Hermite polynomials cannot be expressed as a product of two Hermite polynomials in general, we have not found a way to prove that this is the case for this particular formula that we obtained, and it may very well be the case that this neatly reduces to a product of two(non-orthogonal) Hermite polynomials\cite{Pere1998}. In section \ref{pereversev_discussion}, we discuss that numerical simulation may be the only way to get some sort of a handle on this issue.
\section{Dissipation and heating}
We are able to include the contributions to the beam evolution from hot-plasma dissipation, although the derivation is built on the properties of a cold plasma. Our hot plasma dielectric tensor can be decomposed in the following manner:
\begin{equation}
    \bm{\epsilon}_\mathrm{hot}=\bm{\epsilon}_\mathrm{H}+\bm{\epsilon}_\mathrm{A},
\end{equation}
where \(\bm{\epsilon}_\mathrm{H}\) is the Hermitian part of the plasma dielectric tensor and \(\bm{\epsilon}_\mathrm{A}\) is some anti-Hermitian contribution responsible for heating of resonant particles\cite{van1993dielectric}. Note that other approaches exist for modelling non-Hermitian dielectric tensors\cite{Friedland1980}, but we have not found an approach to integrate such an approach into our model. For simplicity, we assume that \(|\bm{\epsilon}_\mathrm{A}|\sim\frac{\lambda}{L}|\bm{\epsilon}_\mathrm{H}|\). This is to ensure that \(\bm{\epsilon}_\mathrm{A}\) only enters at second order, such that the Hamiltonian is strictly real to lowest order. With this ordering, hot plasma dissipation appears only as a modification to the profile evolution equation, which given as a PDE. Index notation will be utilised as it makes the derivatives explicit without needing to bother oneself with the orderings, it is given by:
\begin{equation}\label{eqn:Profile_Evolution_Dissipative}
    \frac{\partial\mathcal{P}}{\partial\tau}-\mathrm{i}\left(\frac{1}{2}\frac{\partial}{\partial K_\mu} \frac{\partial}{\partial K_\nu} H\frac{\partial}{\partial w_\mu}\frac{\partial}{\partial w_\nu}(\mathcal{P})^{(0)}\right)+w_\mu T_{\mu\nu}\frac{\partial}{\partial w_\nu} \mathcal{P}^{(0)}+\mathrm{i}\hat{e}^*_\mu\epsilon_{A,\mu\nu}(\tau,\bm{w})\hat{e}_\nu\mathcal{P}=0.
\end{equation}
\(\hat{\bm{e}}\) is the polarisation we obtained from solving the cold-plasma dispersion relation. Notice that due to the small length scale of \(\bm{\epsilon}_\mathrm{A}\), it is not only a function of \(\tau\), but one of \(\bm{w}\) as well. This additional width dependence is crucial for calculating power deposition accurately\cite{PARADE2021}. Nonetheless, if it were constant across the width we can easily solve for the amplitude decay, since \(\hat{e}^*_\mu\epsilon_{A,\mu\nu}(\tau,\bm{w})\hat{e}_\nu\) is strictly imaginary. Our amplitude will simply be scaled by \(\exp\left(-\int_0^\tau\mathrm{i}\hat{e}^*_\mu\epsilon_{A,\mu\nu}(\tau')\hat{e}_\nu\mathrm{d}\tau'\right)\), which results in attenuation of our beam due to absorption of energy by electron heating. However, when it depends on \(\bm{w}\), we are no longer able to solve this exactly. Thus, numerically solving eq. \ref{eqn:Profile_Evolution_Dissipative} would still be the only way to approach this. 

\section{Propagation in homogeneous, isotropic media}
A useful testbed for eq. \ref{eqn:Profile_Evolution} and eq. \ref{eqn:Ladder_Operators} is the case of Gauss-Hermite beams in homogeneous media. We will focus on the results in the vacuum case, as propagation in any homogeneous isotropic medium is virtually identical to propagation through vacuum, just with a different refractive index. Thus, in vacuum(or any homogeneous medium), we will use a positively propagating Hamiltonian, \(H=\frac{K^2}{K_0^2}-1\). This is an important consistency check as vacuum propagation of Gauss-Hermite beams is a well established textbook result. Various bases of solutions exist in vacuum, the particular bases of solutions we will focus on are the `elegant' Gauss-Hermite beams, given by:
\begin{equation}
    \bm{E}\propto\hat{\bm{e}}H_n\left(\sqrt{-\mathrm{i}\Psi_{xx}}w_x\right)H_m\left(\sqrt{-\mathrm{i}\Psi_{yy}}w_y\right)\exp\left(\frac{\mathrm{i}}{2}\bm{w}\cdot\bm{\Psi}\cdot\bm{w}\right),
\end{equation}
as well as the traditional Hermite polynomial solution:
\begin{equation}
    \bm{E}\propto\hat{\bm{e}}H_n\left(\sqrt{2\Im(\Psi_{xx})}w_x\right)H_m\left(\sqrt{2\Im(\Psi_{yy})}w_y\right)\exp\left(\frac{\mathrm{i}}{2}\bm{w}\cdot\bm{\Psi}\cdot\bm{w}\right).
\end{equation}
\(H_n(x)\) will only refer to the probabilist's Hermite polynomials, as has been the case throughout our manuscript, since the physicist's Hermite polynomials are simply given by \(H_n(x)_{(\mathrm{physicist})}=\sqrt{2}^nH_n(\sqrt{2}x)_{(\mathrm{probabilist})}\).
It is simple to show that our position, which will always be a straight line distance \textit{z}, will be \(z=\frac{2\tau}{K_0}\). Without loss of generality, we can take the beam to initialise with 0 beam curvature. By solving a simple ODE for the beam envelope, we see that:
\begin{equation}
    \Psi_{ii}=\frac{\Psi_{ii,0}}{1+\Psi_{ii,0}\frac{z}{K_0}}.
\end{equation}
Without loss of generality, we can initialise our beam with 0 curvature, yielding:
\begin{equation}
    \Im(\Psi_{ii})=\frac{\Im(\Psi_{ii})_0}{1+(\Im(\Psi_{ii})_0\frac{z}{K_0})^2}.
\end{equation}

\subsubsection{`Elegant' Hermite polynomial basis\label{eleg_hermite}}
Using the typical coordinates in basic Gaussian beam theory, \(\Im(\Psi_{ii})_0\frac{z}{K_0}=\frac{z}{z_{R,i}}\). We find that the \(\tau\)-derivative of our profile is given by:
\begin{equation}
    \frac{\mathrm{d}}{\mathrm{d}\tau}\left(A(\tau)H_n\left(\sqrt{-\mathrm{i}\Psi_{xx}}w_x\right)H_m\left(\sqrt{-\mathrm{i}\Psi_{yy}}w_y\right) \right)= A\frac{\mathrm{d}\ln(A(\tau))}{\mathrm{d}\tau}H_nH_m+\frac{1}{2}A\bm{w}\cdot\bm{\Psi}^{-1}\frac{\mathrm{d}\bm{\Psi}}{\mathrm{d}\tau}\cdot\bm{\nabla}(H_nH_m).
\end{equation}
The profile evolution equation thus tells us that:
\begin{equation}
    \frac{\mathrm{d}\ln(A(\tau))}{\mathrm{d}\tau}H_nH_m-\mathrm{i}\frac{1}{K_0^2}\left( \nabla^2(H_nH_m)-\bm{w}\cdot(-\mathrm{i}\bm{\Psi})\cdot\bm{\nabla}(H_nH_m) \right)=0,
\end{equation}
but using the fact that \(H_n''-xH_n'=-nH_n\) for the probabilist's Hermite polynomials, we have that:
\begin{equation}
    \ln(A)=C\left(-\frac{n}{2}\ln\left(1+\frac{\mathrm{i}z}{z_{r,x}}\right)-\frac{m}{2}\ln\left(1+\frac{\mathrm{i}z}{z_{r,y}}\right)\right).
\end{equation}
We can split this into a phase and magnitude contribution, from which we find that the magnitude is given by:
\begin{equation}
    A\propto\left(1+\frac{z^2}{z_{r,x}^2}\right)^\frac{-n}{4}\left(1+\frac{z^2}{z_{r,y}^2}\right)^\frac{-m}{4}.
\end{equation}
It is simple to confirm that the overall amplitude modulation is identical to the textbook result. Adding the phase contribution to the Gouy phase(there is no polarisation phase), we then get that the phase of our beam is:
\begin{equation}
    \phi=-\left(\frac{n}{2}\mathrm{arctan}(\frac{z}{z_{R,x}})+\frac{m}{2}\mathrm{arctan}(\frac{z}{z_{R,y}})+\frac{1}{2}\mathrm{arctan}(\frac{z}{z_{R,x}})+\frac{1}{2}\mathrm{arctan}(\frac{z}{z_{R,x}})\right).
\end{equation}
These are both standard textbook results. The magnitude of the amplitude also evolves exactly the same as in textbook Gaussian beam optics.\\
\\
To construct appropriate ladder operators that reduce to Hermite polynomials at the start of propagation, we will heavily reference section \ref{GH_Boundary}. However, we will only present the forward operators as the backward operators in each case are of no use for finding solutions. Firstly, note that we have two possible ways of generating the Hermite polynomials, the first way is by the lowering operator:
\begin{equation}
    \hat{\mathcal{R}}_I=x_I-\partial_{x,I},
\end{equation}
by acting on some constant, typically chosen to be 1. This directly generates the Hermite polynomials in an arbitrary number of dimensions. The second method is using the raising operator:
\begin{equation}
    \hat{\mathcal{L}}_I=-\partial_{x,I},
\end{equation}
by acting on \(f=\exp\left(-\frac{1}{2}x^2\right)\). This generates the Hermite functions in an arbitrary number of dimensions, which are the Hermite polynomials multiplied by \(\exp\left(-\frac{1}{2}|\bm{x}|^2\right)\). If we set \(\bm{\Phi}(\tau=0)=-\bm{\Psi}_{w,0}\), \(\bm{\Psi}_\mathrm{tot}=0\) and remains 0 throughout the evolution in vacuum. Thus, \(\bm{\Phi}(\tau)=-\bm{\Psi}_w(\tau)\) is a valid solution in vacuum. Furthermore, since the diagonal basis of \(\bm{\Psi}\) does not change, we can easily solve for \(\bm{u}_I(\tau)\), and find the following forward operator:
\begin{equation}
    \hat{\mathcal{R}}_I=\sqrt{\frac{\mathrm{i}}{\Psi_{II,0}}}\exp\left(-\ln\left(\frac{\mathrm{i}}{\Psi_{II}}\right)\right)\left(w_I-\frac{\mathrm{i}}{\Psi_{II}}\partial_I\right)=\sqrt{\frac{\Psi_{II}}{\Psi_{II,0}}}\left(\sqrt{\frac{\Psi_{II}}{\mathrm{i}}}w_I-\sqrt{\frac{\mathrm{i}}{\Psi_{II}}}\partial_I\right),
\end{equation}
which acts on a constant function. This operator clearly matches what we need to generate the Gauss-Hermite beams, and has the amplitude correction built into it.\\
\\
Next, we wish to construct our raising operator. If we set \(\bm{\Phi}(\tau=0)=\bm{\Psi}_{w,0}\), \(\bm{\Psi}=0\) and remains 0 throughout the evolution in vacuum. \(\bm{\Phi}=\bm{\Psi}_w\) in this case, and we no longer require a beam envelope as it is built into our profile. Similarly, we can solve for \(\bm{v}_I(\tau)\), obtaining the forward operator:
\begin{equation}
    \hat{\mathcal{L}}_I=-\sqrt{\frac{\mathrm{i}}{\Psi_{II,0}}}\partial_I=-\sqrt{\frac{\Psi_{II}}{\Psi_{II,0}}}\left(\sqrt{\frac{\mathrm{i}}{\Psi_{II}}}\partial_I\right),
\end{equation}
and it acts on \( \exp\left(({\mathrm{i}}/{2})\bm{w}\cdot\bm{\Psi}_w\cdot\bm{w}\right)\). This ladder operator once again has the amplitude evolution built right into it. It is now clear that the complex Hermite polynomials are also given by both our ladder operators, and that the evolution of their profiles are identical for all three approaches. \\
\\
We wish to point out that although the `elegant' Gauss-Hermite beams are biorthogonal under the bilinear pairing inner product(section. \ref{biorthogonality}), this means that the elegant Gauss-Hermite beams do not neatly diagonalise in power, thus it is challenging to analyse the total power across the modes using this basis. An alternative basis of solutions that neatly diagonalises in power is the traditional Gauss-Hermite beam ansatz\cite{pampaloni2004}. 
\subsubsection{Traditional Hermite polynomial basis\label{trad_hermite}}
In this basis, the Hermite polynomials we use are physicist's Hermite polynomials. We can initialise in this basis by setting \(\bm{\Phi}(0)=-2\bm{\Psi}_{w,0}\), and setting \(\bm{\Psi}(0)=\bm{\Psi}_{w,0}\). Thus, \(\bm{\Psi}_\mathrm{tot}(0)=-\bm{\Psi}_{w,0}\). We will only present this solution using the ladder operator approaches, as ladder operator solutions would trivially be solutions to the profile evolution. \\
\\
We will use the exact same ladder operators as before, that is, we are still using the coordinate system \(\bm{x}=\sqrt{\mathrm{i}\bm{\Phi}^\mathrm{(Case\,1)}}\bm{w}\). This has scaled our previous choice of coordinates by \(\sqrt{2}\). Firstly, to obtain \(\bm{\Phi}(\tau)\), we see that:
\begin{equation}
\begin{split}
    \Psi_{\mathrm{tot},ii}&=\frac{-\mathrm{i}W_{i,0}^{-2}}{1-\mathrm{i}\frac{2\tau }{W_{i,0}^2K_0^2}},\\
    \Psi_{ii}&=\frac{\mathrm{i}W_{i,0}^{-2}}{1+\mathrm{i}\frac{2\tau }{W_{i,0}^2K_0^2}},\\
    \therefore \Phi^{(\mathrm{Case\,1})}_{ii}&=-2\mathrm{i}\Im(\Psi_{ii}).
\end{split}
\end{equation}
where we have initialised our coordinates at the waist of the beam. Remarkably, \(\bm{\Phi}\) remains imaginary throughout the evolution if initialised at the beam waist(\(\Psi_{ii,0}=\mathrm{i}W_{i,0}^{-2}\)). Once again, both ladder operator solutions yield the same amplitude modulation. It turns out that there is no amplitude modulation from our ladder operators, but they introduce a phase change for the \textit{nm}-th mode given by:
\begin{equation}
    \phi=n\arctan\left(\frac{2\tau}{W_{x,0}^2K_0^2}\right)+m\arctan\left(\frac{2\tau}{W_{y,0}^2K_0^2}\right),
\end{equation}
which is precisely the textbook phase change we expect. Once again, the overall amplitude and phase modulation is identical to the textbook result.
\\
\\
We can also initialise using the Gaussian to generate our Hermite polynomials, by setting \(\bm{\Phi}(0)=2\bm{\Psi}_{w,0}\), and setting \(\bm{\Psi}(0)=-\bm{\Psi}_{w,0}\). Thus, \(\bm{\Psi}_\mathrm{tot}(0)=\bm{\Psi}_{w,0}\). Importantly, this tells us that \(\bm{\Phi}^\mathrm{(Case\,1)}=-\bm{\Phi}^\mathrm{(Case\,2)}\). We will not demonstrate the procedure this time as it should now be quite straightforward and trivial, but one expectedly yields the same results as before.\\
\\
We reiterate our initial point on the importance of this basis. This basis is diagonal in energy in homogeneous media, as one can prove by substituting the solutions into the bi-orthogonality relation given by eq. \ref{eqn:gaussian_inner_product}. What this means is that when we use this basis of solutions in homogeneous media, the energy contribution to each mode stays constant. This is not the case for the 'elegant' Hermite polynomial basis, so it goes to show that choosing the right basis of solutions at the plasma boundary could have a huge effect on the ease of analysis. \\
\\
In this section, we have also outlined the procedure for initialising with the physicist's Hermite polynomials at the plasma boundary, which serves an important purpose as we are able to assign a certain energy to each mode at the plasma boundary, allowing us to then analyse the energy carried by each mode into the plasma, along with how the energy might mix between modes within the plasma.

\section{Relation to prior work by Pereversev\label{pereversev_discussion}}
To the best our knowledge, prior work by Pereversev\cite{Pere1998} is the most comprehensive and directly comparable attempt at beam tracing of an arbitrary profile. He does so by directly substituting in a Hermite polynomial basis of solutions into the Helmholtz equation, then analysing terms of various orders in the resulting expansion. This approach differs from our approach as it immediately prescribes a basis of solutions, whereas our results allow one to use any suitably chosen basis of solutions or even match boundary conditions exactly, simply by tweaking the ladder operators. An important feature to note is that our evolution equations for the beam envelope is decoupled from the profile evolution, whereas it is coupled to the profile evolution in Pereversev's work. Whilst this may not have the biggest effect numerically, it certainly makes finding analytic solutions much more challenging relative to our approach.\\
\\
One may notice that he obtains similar results to this paper on the whole, but in this section we wish to focus on where our results deviate. We will do this not only by comparing specific arguments and equations in Pereversev's paper, but also by matching our boundary conditions to those matched by Pereversev's solutions. It turns out that our solutions do not seem to match in an obvious way.\\
\\
Pereversev does not treat eqs. \ref{eqn:Amplitude_Evolution_magnitude}, \ref{eqn:Amplitude_Evolution_Phase}, \ref{eqn:Beam_Tracing}, \ref{eqn:Profile_Evolution} separately, but rather considers them together by substituting in a specific basis of solutions. Upon substituting in the ansatz:
\begin{equation}
    \bm{E}_{nm}(\bm{r})=A_{nm}H_n(\sqrt{2}v_1^\mu(\tau) w_\mu)H_m(\sqrt{2}v_2^\mu(\tau) w_\mu)\exp\left(-\frac{1}{2}\left((v_1^\mu w_\mu)^2+(v_2^\mu w_\mu)^2\right)\right)\exp\left(\mathrm{i}S(\bm{r})\right)\hat{\bm{e}},
\end{equation}
we see that in order for this ansatz to be a solution, we first require eq. \ref{eqn:vector_evolution} to be satisfied. Note that eq. 3.2 in Pereversev's paper implies eq. \ref{eqn:vector_evolution}, as \(\frac{\mathrm{d}}{d\tau}(\bm{v}_I\cdot\hat{\bm{g}})=0\), which can be easily proven from eq. 3.2. Explicitly, eq. 3.2 in Pereversev's paper states that:
\begin{equation}
    \frac{\mathrm{d}\bm{v}_I}{\mathrm{d}\tau}=-(\bm{\Psi}\bm{\nabla}_{\bm{K}}\bm{\nabla}_{\bm{K}}(H)+\bm{\nabla}\bm{\nabla}_{\bm{K}}(H))\cdot\bm{v}_I
\end{equation}
Crucially, one also requires that these vectors satisfy \(\bm{v}_1\cdot\bm{\nabla}_{\bm{K}}\bm{\nabla}_{\bm{K}}H\cdot\bm{v}_2=0\), such that we obtain:
\begin{equation}\label{eqn:pereversev_results}
\begin{split}
    &\frac{\mathrm{d}\bm{\Psi}}{\mathrm{d}\tau}+\bm{\Psi}\bm{\nabla}_{\bm{K}}\bm{\nabla}_{\bm{K}}(H)\bm{\Psi}+\bm{\Psi}\bm{\nabla}_{\bm{K}}\bm{\nabla}(H)+\bm{\nabla}\bm{\nabla}_{\bm{K}}(H)\bm{\Psi}\\
    &+\bm{\nabla}\bm{\nabla}(H)=\bm{v}_1\bm{v}_1\cdot\bm{\nabla}_{\bm{K}}\bm{\nabla}_{\bm{K}}(H)\cdot\bm{v}_1\bm{v}_1+\bm{v}_2\bm{v}_2\cdot\bm{\nabla}_{\bm{K}}\bm{\nabla}_{\bm{K}}(H)\cdot\bm{v}_2\bm{v}_2,\\
    &\frac{\mathrm{dln}(A^{(0)})}{\mathrm{d}\tau}+\frac{\mathrm{d}\bm{K}}{\mathrm{d}\tau}\cdot\bm{\nabla}_{\bm{K}} \hat{e}_m\hat{e}^*_m+\hat{e}^*_i\frac{\partial}{\partial K_\mu} D_{im}\partial_\mu\hat{e}_m+\frac{1}{2}\Psi_{\mu\nu}\frac{\partial}{\partial K_\mu}\frac{\partial}{\partial K_\nu} H\\
    &+\mathrm{i}(2n+1)\bm{v}_1\cdot\bm{\nabla}_{\bm{K}}\bm{\nabla}_{\bm{K}}(H)\cdot\bm{v}_1+\mathrm{i}(2m+1)\bm{v}_2\cdot\bm{\nabla}_{\bm{K}}\bm{\nabla}_{\bm{K}}(H)\cdot\bm{v}_2=0.
\end{split}
\end{equation}
These are expectedly identical to Pereversev's equations. We also wish to point out that \(\bm{\Psi}_\mathrm{tot}=\bm{\Psi}+(\bm{v}_1\bm{v}_1+\bm{v}_2\bm{v}_2)\), as defined with Pereversev's evolution equations, obeys eq. \ref{eqn:Beam_Tracing}. Another crucial point is that our \(\bm{v}_I\) and Pereversev's are not the same, as \(\bm{\Psi}\) does not have the same definition. This makes direct comparison of our results very tricky.
\\
\\
In order for \(\bm{v}_1\cdot\bm{\nabla}_{\bm{K}}\bm{\nabla}_{\bm{K}}H\cdot\bm{v}_2=0\) throughout, Pereversev argues that:
\begin{equation}\label{eqn:pereversev_condition}
    \frac{\mathrm{d}^n}{\mathrm{d}\tau^n}(v_1^\mu\frac{\partial}{\partial K_\mu}\frac{\partial}{\partial K_\nu} Hv_2^\nu)|_{\tau=0}=0,
\end{equation}
implying that \(v_1^\mu\frac{\partial}{\partial K_\mu}\frac{\partial}{\partial K_\nu} Hv_2^\nu=0\) throughout the evolution if it is 0 at the boundary. Firstly, note that this solution is a simple product of Hermite polynomials without arguing that they diagonalise \(\bm{\nabla}_{\bm{K}}\bm{\nabla}_{\bm{K}}H\), as if they did they would strictly be orthogonal to each other. However, we argued earlier that they have to be orthogonal in order for a simple product of Hermite polynomials to be the solution in our basis. However, it is still entirely possible that a sum of Hermite polynomial products could neatly reduce to a single product. Since Pereversev's arguments in the Hermite polynomials differ from our arguments, it may still be the case that they are identical. In homogeneous media they are certainly identical.\\
\\
Nonetheless, it is unclear to us how Pereversev proved his result in eq. \ref{eqn:pereversev_condition}, as he only demonstrates it to first order, but the rest does not seem trivially clear to us. If it were true, it suggests the following must be true:
\begin{equation}\label{eqn:gradKgradKH_evolution}
    v_1^\mu v_2^\nu\left(\frac{\mathrm{d}}{\mathrm{d}\tau}\left(\frac{\partial}{\partial K_\mu}\frac{\partial}{\partial K_\nu} H\right)\right)_w=v_1^\mu v_2^\nu\left(T_{w,\mu\rho} \left(\partial K_\rho\frac{\partial}{\partial K_\nu} H\right)_w+T_{w,\nu\rho} \left(\partial K_\rho\frac{\partial}{\partial K_\mu} H\right)_w\right)
\end{equation}
We have not found a way to prove it in Pereversev's paper or in our own attempts. Looking at the terms contracted by \(v_1^\mu v_2^\nu\), we see that there is a real quantity on one side and what seems to generally be a complex quantity on the other side, but it is possible that \(v_1^\mu v_2^\nu\) may conspire in some way such that the equality is satisfied, as they are generally complex vectors in this formulation.\\
\\
Since \(\bm{v}_I\) `diagonalise' \((\frac{\partial}{\partial K_\mu}\frac{\partial}{\partial K_\nu} H)_w\) in vacuum, and remain orthogonal to each other throughout their evolution, Pereversev's solution reproduces the exact same solutions that we obtained earlier in vacuum. This follows naturally from our derivation of Pereversev's beam tracing equations from our profile evolution equation in eq. \ref{eqn:pereversev_results}. Finding a situation in which our solutions do not coincide would prove that eq. \ref{eqn:pereversev_condition} is not satisfied in general. We can only do so in a system where \(\bm{v}_I\) do not diagonalise \((\frac{\partial}{\partial K_\mu}\frac{\partial}{\partial K_\nu} H)_w\), and one good place to look for this is in propagation where the vectors rotate about \(\hat{\bm{g}}\). Only by comparing the predictions of our theory with Pereversev's in that context would one be able to investigate if our theories do in fact give identical results for the Gauss-Hermite boundary conditions. Although analytic work on obtaining the solutions in slab geometry\cite{poli1999paraxial,Tsironis2006} has been done, the vectors do not rotate about \(\hat{\bm{g}}\), nor is \((\frac{\partial}{\partial K_\mu}\frac{\partial}{\partial K_\nu} H)_w\) non-diagonal in the first place. This means that the solutions they obtained will coincide with those obtained by this method, given the same boundary conditions. Unfortunately, we are unable to construct a physical situation where we are able to solve for analytically whilst rotating the vectors, thus this may be a task for numerical comparison.\\
\\
All in all, the consistency between our solutions and Pereversev's is still an open mathematical question. Assuming they are consistent, one additional use of our ladder operator approach is that it enables one to directly model the evolution of any boundary profile in theory, without needing to decompose it in the Hermite polynomial basis. Our ladder operator approach has been demonstrated to directly solve eq. \ref{eqn:Profile_Evolution}, thus we are confident in its legitimacy. It would nonetheless certainly be very interesting theoretically if our two expressions turned out to coincide, as Pereversev's solution is a much simpler form in general for Hermite polynomial profiles.
\section{Implementation of theory}
In this brief section, we discuss the numerical implementation of the theory developed in this paper. One would simply run traditional Gaussian beam tracing\cite{HallChen2022}, and then calculate the evolution of the ladder operators in post. Looking at eq. \ref{eqn:profile_exact_solution}, we see that all we require are $\bm{v}_I(\tau)$ and $\bm{\Phi}(\tau)$. We can easily obtain $\bm{v}_I(\tau)$ from eq. \ref{eqn:vector_evolution} as it is a simple ODE. We will have to run a second beam tracing with the appropriately chosen \(\bm{\Psi}_{w,\mathrm{tot}}=\bm{\Psi}_w+\bm{\Phi}\) in order to obtain $\bm{\Phi}(\tau)$, following the prescription given in eq. \ref{eqn:smart_envelope_choice}. We then substitute these into eq. \ref{eqn:profile_exact_solution} to obtain the beam profile at any point within the plasma, which should computationally be much more lightweight than solving the second-order PDE given in eq. \ref{eqn:Profile_Evolution}.\\
\\
For modelling dissipative effects, one has no choice but to solve eq. \ref{eqn:Profile_Evolution_Dissipative} in regions of dissipation, outside of which our exact solution would hold. Eq. \ref{eqn:Profile_Evolution_Dissipative} may be computationally simpler than solving the full eikonal, however it is still a complicated differential equation to solve in general. However, the ladder operator can be utilised in simple cases to obtain bases of solutions, for example in the case of linear width dependence, which may be the subject of further work.
\subsection{Further Work}
Apart from theoretical work, work must go into computational implementation of our theory. The beam-tracing code of choice would Scotty\cite{HallChen2022}, as our theoretical model will simply reside as an additional module on top of its base beam-tracing capabilities for DBS. This work would hopefully enable Gauss-Hermite DBS analysis in the future, in order to provide more accurate signal analysis for imperfect antenna patterns.\\
\\
In that same vein, we hope to utilise the results of this theory to develop a simplified theory of DBS, and maybe even cross-polarisation scattering(CPS), but with Gauss-Hermite modes -- similar to the approach of previous work in DBS focused on the lowest order Gauss-Hermite beam\cite{HallChen2022, HallChen2022b, HallChen2024}. We hope that this model will be computationally more efficient than evaluating the full volume integral of the reciprocity theorem, such that it is capable of providing real-time synthetic diagnostic analysis. Above all, we believe such a theory would give a clearer account of various factors that can affect the diagnostic signal, enabling accurate extraction of the turbulence spectrum from the signal, beyond just optimising diagnostic design. In order to do this, figuring out discontinuous boundary conditions would also be crucial.
\section{Conclusion}
In this paper, we first derived the beam-tracing equations in a cold plasma for a beam-centred coordinate system. We utilised second-order perturbation theory, obtaining equations which not only provide us with the lowest-order results, but also provide first-order corrections to our beam polarisation. These beam tracing equations govern the evolution of the Gaussian beam envelope, the amplitude of the beam, and the evolution of the beam profile itself.\\
\\
We proceed to obtain a basis of solutions to the profile-evolution equation, for any choice of initial profile. These solutions were obtained using ladder operators, which can be used to set a wide range of polynomial boundary profiles. We focused on constructing ladder operators that generated functions which reduced to the Gauss-Hermite beams at the plasma boundary. 
Finally, we utilised our ladder operators to construct the exact solution to the propagation of a Gauss-Hermite beam through an inhomogeneous cold plasma using eq. \ref{eqn:profile_exact_solution}. Most importantly however, this solution was derived directly from solving eq. \ref{eqn:Profile_Evolution} in full generality, thus we are confident in the accuracy of the result.\\
\\
We finally demonstrated our ladder operators in action for propagation of a Gauss-Hermite beam through homogeneous media, before discussing future research direction, in particular, hopefully applying the results of this paper for improving real-time microwave diagnostics analysis. 
\begin{acknowledgments}
J Ruiz Ruiz was funded by the Engineering and Physical Sciences Research Council (EPSRC), with grant numbers EP/R034737/1 and EP/W026341/1.
\end{acknowledgments}
\bibliography{GHBEAM}

@PREAMBLE{
 "\providecommand{\noopsort}[1]{}" 
 # "\providecommand{\singleletter}[1]{#1}%" 
}

@misc{Leen2016,
      title={Eigenfunctions of the Multidimensional Linear Noise Fokker-Planck Operator via Ladder Operators}, 
      author={Todd K. Leen and Robert Friel and David Nielsen},
      year={2016},
      eprint={1609.01194},
      archivePrefix={arXiv},
      primaryClass={math.CA}
}

@article{HallChen2022,
doi = {10.1088/1361-6587/ac57a1},
year = {2022},
month = {jul},
publisher = {IOP Publishing},
volume = {64},
number = {9},
pages = {095002},
author = {Valerian H Hall-Chen and Felix I Parra and Jon C Hillesheim},
title = {Beam model of Doppler backscattering},
journal = {Plasma Physics and Controlled Fusion}
}

@article{Poli2001,
title = {TORBEAM, a beam tracing code for electron-cyclotron waves in tokamak plasmas},
journal = {Computer Physics Communications},
volume = {136},
number = {1},
pages = {90-104},
year = {2001},
issn = {0010-4655},
doi = {https://doi.org/10.1016/S0010-4655(01)00146-1},

author = {E. Poli and A.G. Peeters and G.V. Pereverzev}
}

@article{Pere1998,
    author = {Pereverzev, G. V.},
    title = {Beam tracing in inhomogeneous anisotropic plasmas},
    journal = {Physics of Plasmas},
    volume = {5},
    number = {10},
    pages = {3529-3541},
    year = {1998},
    month = {10},
    issn = {1070-664X},
    doi = {10.1063/1.873070}
}

@article{Erckmann1994,
doi = {10.1088/0741-3335/36/12/001},
year = {1994},
month = {dec},
publisher = {},
volume = {36},
number = {12},
pages = {1869},
author = {V Erckmann and U Gasparino},
title = {Electron cyclotron resonance heating and current drive in toroidal fusion plasmas},
journal = {Plasma Physics and Controlled Fusion}
}

@misc{pampaloni2004,
      title={Gaussian, Hermite-Gaussian, and Laguerre-Gaussian beams: A primer}, 
      author={Francesco Pampaloni and Joerg Enderlein},
      year={2004},
      eprint={physics/0410021},
      archivePrefix={arXiv},
      primaryClass={physics.optics}
}

@article{PARADE2021,
    author = {Yanagihara, K. and Dodin, I. Y. and Kubo, S.},
    title = {Quasioptical modeling of wave beams with and without mode conversion. IV. Numerical simulations of waves in dissipative media},
    journal = {Physics of Plasmas},
    volume = {28},
    number = {12},
    pages = {122102},
    year = {2021},
    month = {12},
    issn = {1070-664X},
    doi = {10.1063/5.0057345}
}

@article{Poli2018,
title = {TORBEAM 2.0, a paraxial beam tracing code for electron-cyclotron beams in fusion plasmas for extended physics applications},
journal = {Computer Physics Communications},
volume = {225},
pages = {36-46},
year = {2018},
issn = {0010-4655},
doi = {10.1016/j.cpc.2017.12.018},
author = {E. Poli and A. Bock and M. Lochbrunner and O. Maj and M. Reich and A. Snicker and A. Stegmeir and F. Volpe and N. Bertelli and R. Bilato and G.D. Conway and D. Farina and F. Felici and L. Figini and R. Fischer and C. Galperti and T. Happel and Y.R. Lin-Liu and N.B. Marushchenko and U. Mszanowski and F.M. Poli and J. Stober and E. Westerhof and R. Zille and A.G. Peeters and G.V. Pereverzev}
}

@inproceedings{weber2015,
  title={Scattering of diffracting beams of electron cyclotron waves by random density fluctuations in inhomogeneous plasmas},
  author={Weber, Hannes and Maj, Omar and Poli, Emanuele},
  booktitle={EPJ Web of Conferences},
  volume={87},
  pages={01002},
  year={2015},
  organization={EDP Sciences}
}

@article{Snicker2018,
doi = {10.1088/1741-4326/aa8d07},
year = {2017},
month = {nov},
publisher = {IOP Publishing},
volume = {58},
number = {1},
pages = {016002},
author = {Snicker, A. and Poli, E. and Maj, O. and Guidi, L. and Köhn, A. and Weber, H. and Conway, G. and Henderson, M. and Saibene, G.},
title = {The effect of density fluctuations on electron cyclotron beam broadening and implications for ITER},
journal = {Nuclear Fusion}
}

@article{Friedland1980,
  title = {Geometric optics in plasmas characterized by non-Hermitian dielectric tensors},
  author = {Friedland, L. and Bernstein, I. B.},
  journal = {Phys. Rev. A},
  volume = {22},
  issue = {4},
  pages = {1680--1685},
  numpages = {0},
  year = {1980},
  month = {Oct},
  publisher = {American Physical Society},
  doi = {10.1103/PhysRevA.22.1680}
}

@article{LaHaye2009,
doi = {10.1088/0029-5515/49/4/045005},
year = {2009},
month = {mar},
publisher = {},
volume = {49},
number = {4},
pages = {045005},
author = {La Haye, R.J. and Isayama, A. and Maraschek, M.},
title = {Prospects for stabilization of neoclassical tearing modes by electron cyclotron current drive in ITER},
journal = {Nuclear Fusion}
}

@misc{yeoh2025,
      title={Conceptual study on using Doppler backscattering to measure magnetic pitch angle in tokamak plasmas}, 
      author={AK Yeoh and VH Hall-Chen and QT Pratt and BS Victor and J Damba and TL Rhodes and NA Crocker and KR Fong and JC Hillesheim and FI Parra and J Ruiz Ruiz},
      year={2025},
      eprint={2502.19061},
      archivePrefix={arXiv},
      primaryClass={physics.plasm-ph} 
}

@misc{liang2025,
      title={Conceptual design of a Doppler Backscattering diagnostic for the EXL-50U spherical tokamak}, 
      author={Ying Hao Matthew Liang and Valerian Hongjie Hall-Chen and Terry L. Rhodes and Yumin Wang and Yihang Zhao},
      year={2025},
      eprint={2509.18532},
      archivePrefix={arXiv},
      primaryClass={physics.plasm-ph}
}

@article{RuizRuiz2025, title={Beam focusing and consequences for Doppler backscattering measurements}, volume={91}, DOI={10.1017/S0022377825000170}, number={2}, journal={Journal of Plasma Physics}, author={Ruiz Ruiz, J. and Parra, F.I. and Hall-Chen, V.H. and Belrhali, N. and Giroud, C. and Hillesheim, J.C. and Lopez, N.A.}, year={2025}}

@article{Hillesheim2009,
    author = {Hillesheim, J. C. and Peebles, W. A. and Rhodes, T. L. and Schmitz, L. and Carter, T. A. and Gourdain, P.-A. and Wang, G.},
    title = {A multichannel, frequency-modulated, tunable Doppler backscattering and reflectometry system},
    journal = {Review of Scientific Instruments},
    volume = {80},
    number = {8},
    pages = {083507},
    year = {2009},
    month = {08},
    issn = {0034-6748},
    doi = {10.1063/1.3205449}
}

@article{Hennequin2006,
doi = {10.1088/0029-5515/46/9/S12},
url = {https://doi.org/10.1088/0029-5515/46/9/S12},
year = {2006},
month = {aug},
publisher = {},
volume = {46},
number = {9},
pages = {S771},
author = {Hennequin, P. and Honoré, C. and Truc, A. and Quéméneur, A. and Fenzi-Bonizec, C. and Bourdelle, C. and Garbet, X. and Hoang, G.T. and the Tore Supra team},
title = {Fluctuation spectra and velocity profile from Doppler backscattering on Tore Supra},
journal = {Nuclear Fusion}
}

@misc{HallChen2024,
      title={Effect of mismatch on Doppler backscattering in MAST and MAST-U plasmas}, 
      author={Valerian H. Hall-Chen and Felix I. Parra and Jon C. Hillesheim and Juan Ruiz Ruiz and Neal A. Crocker and Peng Shi and Hong Son Chu and Simon J. Freethy and Lucy A. Kogan and William A. Peebles and Quinn T. Pratt and Terry L. Rhodes and Kevin Ronald and Rory Scannell and David C. Speirs and Stephen Storment and Jonathan Trisno},
      year={2024},
      eprint={2211.17141},
      archivePrefix={arXiv},
      primaryClass={physics.plasm-ph} 
}

@article{HallChen2022b,
    author = {Hall-Chen, V. H. and Damba, J. and Parra, F. I. and Pratt, Q. T. and Michael, C. A. and Peng, S. and Rhodes, T. L. and Crocker, N. A. and Hillesheim, J. C. and Hong, R. and Ni, S. and Peebles, W. A. and Png, C. E. and Ruiz Ruiz, J.},
    title = {Validating and optimizing mismatch tolerance of Doppler backscattering measurements with the beam model (invited)},
    journal = {Review of Scientific Instruments},
    volume = {93},
    number = {10},
    pages = {103536},
    year = {2022},
    month = {10},
    issn = {0034-6748},
    doi = {10.1063/5.0101805}
}

@article{gusakov2004,
  title={Spatial and wavenumber resolution of Doppler reflectometry},
  author={Gusakov, EZ and Surkov, AV},
  journal={Plasma physics and controlled fusion},
  volume={46},
  number={7},
  pages={1143},
  year={2004},
  publisher={IOP Publishing}
}

@article{Conway2004,
doi = {10.1088/0741-3335/46/6/003},
year = {2004},
month = {apr},
publisher = {},
volume = {46},
number = {6},
pages = {951},
author = {G D Conway and J Schirmer and S Klenge and W Suttrop and E Holzhauer and the ASDEX Upgrade Team},
title = {Plasma rotation profile measurements using Doppler reflectometry},
journal = {Plasma Physics and Controlled Fusion}
}

@inproceedings{colas1996,
  title={Parametric analysis of internal magnetic fluctuations in the Tore Supra tokamak},
  author={Colas, L and Zou, XL and Paume, M and Chareau, JM and Guiziou, L and Hoang, GT and Gresillon, D},
  booktitle={Abstracts of the 23rd European physical society conference on controlled fusion and plasma physics},
  number={INIS-UA--043},
  pages={112--112},
  year={1996}
}

@inproceedings{stepanov1998,
  title={Cross-Polarisation Scattering Experiments on the RTP Tokamak},
  author={Stepanov, A Yu and others},
  booktitle={Proc. 24th Int. Conf. on Plasma Phys. Control. Fusion, Prague},
  year={1998}
}

@article{sysoeva2015,
  title={Electron cyclotron resonance heating beam broadening in the edge turbulent plasma of fusion machines},
  author={Sysoeva, EV and Da Silva, F and Gusakov, EZ and Heuraux, St{\'e}phane and Popov, A Yu},
  journal={Nuclear Fusion},
  volume={55},
  number={3},
  pages={033016},
  year={2015},
  publisher={IOP Publishing}
}

@article{wolf2018,
  title={Electron-cyclotron-resonance heating in Wendelstein 7-X: A versatile heating and current-drive method and a tool for in-depth physics studies},
  author={Wolf, RC and Bozhenkov, S and Dinklage, A and Fuchert, G and Kazakov, Ye O and Laqua, HP and Marsen, S and Marushchenko, NB and Stange, T and Zanini, M and others},
  journal={Plasma Physics and Controlled Fusion},
  volume={61},
  number={1},
  pages={014037},
  year={2018},
  publisher={IOP Publishing}
}

@Inbook{Risken1996,
author="Risken, Hannes",
title="Fokker-Planck Equation for Several Variables; Methods of Solution",
bookTitle="The Fokker-Planck Equation: Methods of Solution and Applications",
year="1996",
publisher="Springer Berlin Heidelberg",
address="Berlin, Heidelberg",
pages="133--162",
isbn="978-3-642-61544-3",
doi="10.1007/978-3-642-61544-3_6"
}

@article{Vatiwutipong2019,
    author = {Vatiwutipong, P and Phewchean, N},
    title = {Alternative way to derive the distribution of the multivariate Ornstein–Uhlenbeck process},
    journal = {Advances in Difference Equations},
    year = {2019},
    doi ={10.1186/s13662-019-2214-1}
}

@article{smirnov2001genray,
  title={The GENRAY ray tracing code},
  author={Smirnov, AP and Harvey, RW},
  journal={CompX Report CompX-2000-01},
  year={2001}
}

@article{Esterkin1996,
doi = {10.1088/0029-5515/36/11/I05},
year = {1996},
month = {nov},
publisher = {},
volume = {36},
number = {11},
pages = {1501},
author = {A.R. Esterkin and A.D. Piliya},
title = {Fast ray tracing code for LHCD simulations},
journal = {Nuclear Fusion}
}

@article{xie2022boray,
  title={BORAY: A ray tracing code for various magnetized plasma configurations},
  author={Xie, Huasheng and Banerjee, Debabrata and Bai, Yukun and Zhao, Hanyue and Li, Jingchun},
  journal={Computer Physics Communications},
  volume={276},
  pages={108363},
  year={2022},
  publisher={Elsevier}
}

@article{Tsironis2006,
    author = {Tsironis, Christos and Poli, Emanuele and Pereverzev, Grigory V.},
    title = {Beam tracing description of non-Gaussian wave beams},
    journal = {Physics of Plasmas},
    volume = {13},
    number = {11},
    pages = {113304},
    year = {2006},
    month = {11},
    issn = {1070-664X},
    doi = {10.1063/1.2390686}
}

@article{poli1999paraxial,
  title={Paraxial Gaussian wave beam propagation in an anisotropic inhomogeneous plasma},
  author={Poli, E and Pereverzev, GV and Peeters, AG},
  journal={Physics of Plasmas},
  volume={6},
  number={1},
  pages={5--11},
  year={1999},
  publisher={American Institute of Physics}
}

@misc{Wikarta2025, title={Beam tracing for stellarators}, publisher={Nanyang Technological University}, author={Wikarta, Eduard}, year={2025}}

@misc{maheshwari2014,
      title={Properties of Tensor Hermite Polynomials}, 
      author={Parul Maheshwari and Gautam Mukhopadhyay and Siddhartha SenGupta},
      year={2014},
      eprint={1411.7398},
      archivePrefix={arXiv},
      primaryClass={math-ph}
}

@article{grad1949,
  title={Note on N-dimensional hermite polynomials},
  author={Grad, Harold},
  journal={Communications on Pure and Applied Mathematics},
  volume={2},
  number={4},
  pages={325--330},
  year={1949},
  publisher={Wiley Online Library}
}

@article{zhu2015polarimetry,
  title={Reconstruction of the density profile for the EAST tokamak based on polarimeter/interferometer and microwave reflectometer systems},
  author={Zhu, Xiang and Zeng, Long and Liu, Haiqing and Jie, Yinxian and Zhang, Shoubiao and Hu, Jiansheng and Gao, Xiang},
  journal={Plasma Science and Technology},
  volume={17},
  number={9},
  pages={733},
  year={2015},
  publisher={IOP Publishing}
}

@article{smith2008polarimetry,
  title={Nonperturbative measurement of the local magnetic field using pulsed polarimetry for fusion reactor conditions},
  author={Smith, Roger J},
  journal={Review of Scientific Instruments},
  volume={79},
  number={10},
  year={2008},
  publisher={AIP Publishing}
}

@article{Kohn2025review,
    author = {Köhn-Seemann, Alf and B. Morales, Rennan},
    title = {From electron cyclotron emission and reflectometry to microwave imaging diagnostics in fusion plasmas: Progress and perspectives},
    journal = {Physics of Plasmas},
    volume = {32},
    number = {6},
    pages = {060502},
    year = {2025},
    month = {06},
    issn = {1070-664X},
    doi = {10.1063/5.0259713}
}

@Article{Goncalves2023engineering,
AUTHOR = {Gonçalves, Bruno and Varela, Paulo and Silva, António and Silva, Filipe and Santos, Jorge and Ricardo, Emanuel and Vale, Alberto and Luís, Raúl and Nietiadi, Yohanes and Malaquias, Artur and Belo, Jorge and Dias, José and Ferreira, Jorge and Franke, Thomas and Biel, Wolfgang and Heuraux, Stéphane and Ribeiro, Tiago and De Masi, Gianluca and Tudisco, Onofrio and Cavazzana, Roberto and Marchiori, Giuseppe and D’Arcangelo, Ocleto},
TITLE = {Advances, Challenges, and Future Perspectives of Microwave Reflectometry for Plasma Position and Shape Control on Future Nuclear Fusion Devices},
JOURNAL = {Sensors},
VOLUME = {23},
YEAR = {2023},
NUMBER = {8},
ARTICLE-NUMBER = {3926},
ISSN = {1424-8220},
DOI = {10.3390/s23083926}
}

@article{chen2018polarimetry,
  title={A Faraday-effect polarimeter for fast magnetic dynamics measurement on DIII-D},
  author={Chen, J and Ding, WX and Brower, DL and Finkenthal, D and Boivin, R},
  journal={Review of Scientific Instruments},
  volume={89},
  number={10},
  year={2018},
  publisher={AIP Publishing}
}

@article{bornatici1983ece,
  title={Electron cyclotron emission and absorption in fusion plasmas},
  author={Bornatici, M and Cano, R and De Barbieri, O and Engelmann, F},
  journal={Nuclear Fusion},
  volume={23},
  number={9},
  pages={1153},
  year={1983},
  publisher={IOP Publishing}
}

@article{liu2018ece,
  title={Overview of the electron cyclotron emission measurements on EAST},
  author={Liu, Yong and Zhao, Hailin and Zhou, Tianfu and Liu, Xiang and Zhu, Zeying and Han, Xiang and Schmuck, Stefan and Fessey, John and Trimble, Paul and Domier, CW and others},
  journal={Fusion Engineering and Design},
  volume={136},
  pages={72--75},
  year={2018},
  publisher={Elsevier}
}

@article{classen2010ece,
  title={2D electron cyclotron emission imaging at ASDEX Upgrade},
  author={Classen, IGJ and Boom, JE and Suttrop, W and Schmid, E and Tobias, B and Domier, CW and Luhmann, NC and Donn{\'e}, AJH and Jaspers, RJE and De Vries, PC and others},
  journal={Review of Scientific Instruments},
  volume={81},
  number={10},
  year={2010},
  publisher={AIP Publishing}
}

@article{Lin2024reflectometry,
    author = {Lin, Y. and Nikolaeva, V. and Hachmeister, D. and Kowalski, E. and Reinke, M. L.},
    title = {Edge scanning reflectometry for density profile measurement on the SPARC tokamak},
    journal = {Review of Scientific Instruments},
    volume = {95},
    number = {8},
    pages = {083540},
    year = {2024},
    month = {08},
    issn = {0034-6748},
    doi = {10.1063/5.0219533}
}

@article{van1993dielectric,
  title={The dielectric tensor for an arbitrary distribution function},
  author={Van Eester, Dirk},
  journal={Plasma physics and controlled fusion},
  volume={35},
  number={4},
  pages={441},
  year={1993},
  publisher={IOP Publishing}
}

@phdthesis{lopez2022metaplectic,
  title={Metaplectic geometrical optics},
  author={Lopez, Nicolas Alexander},
  year={2022},
  school={Princeton University}
}

@article{donnelly2021steepest,
  title={Steepest-descent algorithm for simulating plasma-wave caustics via metaplectic geometrical optics},
  author={Donnelly, Sean M and Lopez, Nicolas A and Dodin, IY},
  journal={Physical Review E},
  volume={104},
  number={2},
  pages={025304},
  year={2021},
  publisher={APS}
}

@article{hojlund2024metademo,
  title={Demonstration of metaplectic geometrical optics for reduced modeling of plasma waves},
  author={H{\o}jlund Marholt, Rune and Senstius, Mads Givskov and Nielsen, Stefan Kragh},
  journal={Physical Review E},
  volume={110},
  number={2},
  pages={025208},
  year={2024},
  publisher={APS}
}

@article{lopez2024,
  title={Regarding the extension of metaplectic geometrical optics to modeling evanescent waves in ray-tracing codes},
  author={Lopez, NA and H{\o}jlund, R and Senstius, MG},
  journal={Physics of Plasmas},
  volume={31},
  number={8},
  year={2024},
  publisher={AIP Publishing}
}
\appendix

\section{Derivation of beam tracing equations\label{beam_tracing_derivation}}
In this appendix, we will utilise Einstein summation convention throughout. As per convention, the symbol `$\partial_\mu$' refers to the partial derivative with respect to a given coordinate chart $r_\mu$, which in this case is any chosen coordinate chart for our three-dimensional Euclidean space. `$\frac{\partial}{\partial K_\mu}$' refers to the partial derivative of a function with respect to $K_\mu$. This is to facilitate simpler manipulation of the numerous contractions and identities we will be introducing in the proceeding appendix. At the end, the results we obtain can easily be expressed in tensor notation.
\subsection{0th order expansion}
To lowest order, the Helmholtz equation would be:
\begin{equation}
    \mathcal{P}A^{(0)}\Big((\delta_{il}\delta_{jm}-\delta_{ij}\delta_{ml})\partial_m\phi^{(0)}\partial_l\phi^{(0)}+\epsilon_{ij}\Big)\hat{e}_j=0
\end{equation}
This provides us with the dispersion relation for our central ray.
\begin{equation}\label{eqn:dispersion relation_appendix}
    D_{ij}\hat{e}_j=H\hat{e}_i=0,
\end{equation}
where
\begin{equation}
    D_{ij}(\bm{K},\bm{r})=(\delta_{il}\delta_{jm}-\delta_{ij}\delta_{ml})K_mK_l+\epsilon_{ij}.
\end{equation}
We included \textit{H} to reveal the Hamiltonian character of our dispersion relation later on.

\subsection{1st order expansion}
The first order terms of our Helmholtz equation are:
\begin{equation}
\begin{split}
    (\delta_{il}\delta_{jm}-\delta_{im}\delta_{jl})\Big(-\partial_j\phi^{(0)}\partial_l\phi^{(0)} A_m^{(1)}-(\partial_j\phi^{(0)}\partial_l\phi^{(1)}+\partial_l\phi^{(0)}\partial_j\phi^{(1)})A_m^{(0)}&\\
    +\mathrm{i}(\partial_j\phi^{(0)}(\partial_l A_m)^{(0)}+\partial_l\phi^{(0)}(\partial_j A_m)^{(0)})\Big)&=\epsilon_{ij}A_j^{(1)}+w_k\partial_k\epsilon_{ij}A_j^{(0)}  
\end{split}
\end{equation}
By contracting with the conjugate polarisation, \(\hat{e}^*_i\), we can eliminate the term involving the first order correction to the amplitude, as it is contracted by \(\hat{e}^*_iD_{ij}=0\). We will define \((\bm{\nabla}_{\bm{K}})_i=\frac{\partial}{\partial K_i}\). Utilising the fact that:
\begin{equation}
    \partial_lD_{ij}=\partial_l\epsilon_{ij},
\end{equation}
and:
\begin{equation}
    \frac{\partial}{\partial K_\mu} D_{im}=\delta_{i\mu}K_m+\delta_{\mu m}K_i-2\delta_{im}K_\mu=(\delta_{il}\delta_{jm}-\delta_{im}\delta_{jl})(\delta_{j\mu}\delta_{l\nu}+\delta_{j\nu}\delta_{l\mu})K_\nu,
\end{equation}
along with \ref{eqn:dispersion relation_appendix}, we can manipulate our equation into the following form:
\begin{equation}
    \Bigg(\frac{\partial}{\partial K_l}(H)\Big(\Psi_{lk}w_k-\mathrm{i}(\partial_l(\ln\mathcal{P}))^{(0)}\Big)+w_l\partial_l(H)\Bigg)A^{(0)}=0 . 
\end{equation}
In order for this to be satisfied, since \(\bm{\Psi}_w\) is complex and \((\partial_l(\mathcal{P}))^{(0)}\) lies along \(\bm{w}\), \(\bm{\nabla}_KH\propto\bm{g}\). The equation then reduces to:
\begin{equation}
    w_l\left(\frac{1}{g}\frac{\mathrm{d}K_l}{\mathrm{d}\tau}\hat{g}_\mu\frac{\partial}{\partial K_\mu}H+\partial_lH\right)=0.
\end{equation}
We will simply define:
\begin{equation}
    \frac{\partial}{\partial K_i}H=g_i.
\end{equation}
This merely sets the parametrisation \(\tau\). In theory we could choose any other choice of parametrisation so long as \(\bm{\nabla}_KH\propto\bm{g}\), but there is no physical difference to this. With this choice, we also see that:
\begin{equation}
    \left(\frac{\mathrm{d}K_l}{\mathrm{d}\tau}+\partial_lH\right)_w=0.
\end{equation}
To get the component along \(\hat{\bm{g}}\), note that along the central ray,
\begin{equation}
    g_i\partial_iH+\frac{\mathrm{d}K_l}{\mathrm{d}\tau}\frac{\partial}{\partial K_i}H=\frac{\mathrm{d}H}{\mathrm{d}\tau}=0.
\end{equation}
Thus we arrive at the ray tracing equations:
\begin{equation}\label{eqn:ray-tracing_appendix}
\begin{split}
    \frac{\partial}{\partial K_i}H&=\frac{\mathrm{d}q_i}{\mathrm{d}\tau},\\
    \partial_iH&=-\frac{\mathrm{d}K_i}{\mathrm{d}\tau}.
\end{split}
\end{equation}
We can now substitute this result back into our first order expansion, in order to solve for the first order correction to the amplitude function. This is necessary to perform our second order expansion later on. Our first order terms are:
\begin{equation}
    w_k\partial_kD_{ij}\hat{e}_j\mathcal{P}A^{(0)}+\mathcal{P}A^{(0)}\frac{\partial}{\partial K_l}D_{im}\Big(\Psi_{lk}w_k-\mathrm{i}(\partial_l(\ln\mathcal{P}))^{(0)}\Big)\hat{e}_m=-D_{ij}A_j^{(1)}.  
\end{equation}
Using the fact that:
\begin{equation}
\begin{split}
\partial_k(D_{ij}\hat{e}_j)&=\partial_kD_{ij}\hat{e}_j+D_{ij}\partial_k\hat{e}_j,\\
&=\partial_kH\hat{e}_i,  
\end{split}
\end{equation}
for any derivative operator, we can substitute this into our  equation to obtain:
\begin{equation}
    w_k(\partial_kH\hat{e}_i-D_{ij}\partial_k\hat{e}_j)\mathcal{P}A^{(0)}+\mathcal{P}A^{(0)}\left(\frac{\partial}{\partial K_l}H\hat{e}_i-D_{ij}\frac{\partial}{\partial K_l}\hat{e}_j\right)\Big(\Psi_{lk}w_k-\mathrm{i}(\partial_l(\ln\mathcal{P}))^{(0)}\Big)=-D_{ij}A_j^{(1)}.  
\end{equation}
We can then utilise our ray tracing equations, eq. \ref{eqn:ray-tracing_appendix}, to further simplify this, obtaining:
\begin{equation}
    w_k\left(-\frac{\mathrm{d}K_k}{\mathrm{d}\tau}\hat{e}_i-D_{ij}\partial_k\hat{e}_j\right)\mathcal{P}A^{(0)}+\mathcal{P}A^{(0)}\left(g_l\hat{e}_i-D_{ij}\frac{\partial}{\partial K_l}\hat{e}_j\right)\Big(\Psi_{lk}w_k-\mathrm{i}(\partial_l(\ln\mathcal{P}))^{(0)}\Big)=-D_{ij}A_j^{(1)},  
\end{equation}
which further simplifies to:
\begin{equation}
    D_{ij}\mathcal{P}A^{(0)}\left(w_k\partial_k\hat{e}_j+\Big(\Psi_{lk}w_k-\mathrm{i}(\partial_l(\ln\mathcal{P}))^{(0)}\Big)\frac{\partial}{\partial K_l}\hat{e}_j\right)=D_{ij}A_j^{(1)}.
\end{equation}
Note that this equation only constrains the components of \(\bm{A}^{(1)}\) perpendicular to \(\hat{\bm{e}}\). We are somewhat free to add any vector proportional to \(\hat{\bm{e}}\) to \(\bm{A}^{(1)}\), but given that it is a correction, we can simply choose to absorb any such vector into \(\bm{A}^{(0)}\) by changing the definition of \(\hat{\bm{e}}\). We can add any imaginary vector proportional to \(\hat{\bm{e}}\) to \(\partial_i\hat{\bm{e}}\) simply by changing the phase function, that is, by changing \(\hat{\bm{e}}\rightarrow\hat{\bm{e}}\exp(\mathrm{i}\alpha)\):
\begin{equation}
    \partial_i\hat{\bm{e}}\rightarrow\partial_i\hat{\bm{e}}+\mathrm{i}\partial_i\alpha\hat{\bm{e}}.
\end{equation}
Noting that since \(\hat{\bm{e}}^*\hat{\bm{e}}=1\), \(\hat{\bm{e}}^*\partial_i\hat{\bm{e}}\) must be purely imaginary. Therefore, we could choose an appropriate choice of phase function such that all \(\partial_i\hat{\bm{e}}\) are orthogonal to \(\hat{\bm{e}}\). The definition for  \(\hat{\bm{e}}\) is ultimately inconsequential, and \(\bm{A}^{(1)}\) can have a component along \(\bm{A}^{(0)}\). What we have shown is that any components of \(\partial_i\hat{\bm{e}}\) along \(\hat{\bm{e}}\) are only affected by phase information, which is not physically significant, and is a quantity freely defined by us. The only physical significance lies in \(\bm{A}^{(1)}\) perpendicular to \(\bm{A}^{(0)}\). With that in mind, we can take \(\bm{A}^{(1)}\) to be:
\begin{equation}
\mathcal{P}A^{(0)}\left(w_k\partial_k\hat{e}_j+\Big(\Psi_{lk}w_k-\mathrm{i}(\partial_l(\ln\mathcal{P}))^{(0)}\Big)\frac{\partial}{\partial K_l}\hat{e}_j\right)=A_j^{(1)}.
\end{equation}
In other words, we have obtained the first order correction to the polarisation:
\begin{equation}\label{eqn:A1}
\left(w_k\partial_k+\Big(\Psi_{lk}w_k-\mathrm{i}(\partial_l(\ln\mathcal{P}))^{(0)}\Big)\frac{\partial}{\partial K_l}\right)\hat{e}^{(0)}_j=e_j^{(1)}.
\end{equation}

\subsection{2nd order expansion}
The second order terms of the Helmholtz equation are:
\begin{equation}
\begin{split}
    (\delta_{il}\delta_{jm}-\delta_{im}\delta_{jl})\Big(-\partial_j\phi^{(0)}\partial_l\phi^{(0)} A_m^{(2)}&\\
    -(\partial_j\phi^{(0)}\partial_l\phi^{(1)}+\partial_l\phi^{(0)}\partial_j\phi^{(1)})A_m^{(1)}&\\
    -(\partial_j\phi^{(1)}\partial_l\phi^{(1)})A_m^{(0)}&\\ 
    -(\partial_j\phi^{(2)}\partial_l\phi^{(0)}+\partial_l\phi^{(2)}\partial_j\phi^{(0)})A_m^{(0)}&\\
    +\mathrm{i}(\partial_j\phi^{(1)}(\partial_l A_m)^{(0)}+\partial_l\phi^{(1)}(\partial_j A_m)^{(0)})&\\
    +\mathrm{i}(\partial_j\phi^{(0)}(\partial_l A_m)^{(1)}+\partial_l\phi^{(0)}(\partial_j A_m)^{(1)})&\\
     +\mathrm{i}\partial_j\partial_l\phi^{(0)} A_m +\partial_j\partial_l (A_m)^{(0)} \Big)&=\epsilon_{ij}A_j^{(2)}+w_k\partial_k\epsilon_{ij}A_j^{(1)}+  \frac{1}{2}w_kw_l\partial_k\partial_l\epsilon_{ij}A_j^{(0)}.
\end{split}
\end{equation}
We now require \((\partial_iA_m)^{(1)}\). For ease of book-keeping, we will define:
\begin{equation}
    (\alpha^{(1)}_j)_\mathrm{original}=w_k\left(\Psi_{kl}\frac{\partial}{\partial K_l}+\partial_k\right)\hat{e}_jA^{(0)},
\end{equation}
which is akin to \(\bm{A}^{(1)}\) in a previous paper on Gaussian beam tracing\cite{HallChen2022}. Then, \((\partial_iA_m)^{(1)}\) is given by:
\begin{equation}
    (\partial_jA_m)^{(1)}=\partial_j(\mathcal{P})^{(1)}A^{(0)}_m+\mathcal{P}(\partial_j(A^{(0)}_m+({\alpha_m^{(1)}})_\mathrm{original}))^{(0)}-\mathrm{i}\left(\frac{\partial}{\partial K_l} \hat{e}_m \partial_j\partial_l(\mathcal{P})^{(0)}\right)\alpha_m^{(0)},
\end{equation}
where:
\begin{equation}
    \partial_\nu(A^{(0)}_m+({\alpha_m^{(1)}})_\mathrm{original}))^{(0)}=\frac{\hat{g}_\nu}{g}\left( \frac{\mathrm{d}A^{(0)}}{\mathrm{d}\tau}\hat{e}_m+A^{(0)}\frac{\mathrm{d}\hat{e}_m}{\mathrm{d}\tau} \right)+\left(\delta_{\nu k}-\hat{g}_\nu\hat{g}_k)(\Psi_{k l}\frac{\partial}{\partial K_l}+\partial_k\right)\hat{e}_mA^{(0)}.
\end{equation}
This is due to the dependence of \( ({\alpha_m^{(1)}})_\mathrm{original} \) on \(\bm{w}\), whereas \(A^{(0)}\) only depends on \(\tau\). Substituting this into our second-order expansion of the Helmholtz equation, we have:
\begin{equation}\label{eqn:second_order_helmholtz}
\begin{split}
    \hat{e}^*_i\Bigg(&(\delta_{il}\delta_{jm}-\delta_{im}\delta_{jl})(\delta_{j\mu}\delta_{l\nu}+\delta_{j\nu}\delta_{l\mu})\\
    &\Big(-K_\mu\Psi_{\nu k}w_k\Big(\overbrace{-\mathrm{i}\frac{\partial}{\partial K_l}\hat{e}_m\partial_l(\mathcal{P})^{(0)}A^{(0)}}^\mathrm{1}+\mathcal{P}(\alpha^{(1)}_m)_\mathrm{original}\Big)\\
    &-(\frac{1}{2}\Psi_{\mu k}w_k\Psi_{\nu j}w_j+K_\mu\partial_\nu\phi^{(2)})\hat{e}_m\mathcal{P}A^{(0)}\\
    &\overbrace{+\mathrm{i}(\Psi_{\mu k}w_k\partial_\nu(\mathcal{P})^{(0)}\hat{e}_m)A^{(0)}}^\mathrm{1}\\
    &+\mathrm{i}K_\mu\Big(\mathcal{P}\partial_\nu(A^{(0)}_m+(\alpha^{(1)}_m)_\mathrm{original})+\overbrace{\partial_\nu \mathcal{P}^{(1)}A^{(0)}_m}^\mathrm{4}+\overbrace{\partial_\nu \mathcal{P}^{(0)}(\alpha^{(1)}_m)_\mathrm{original}}^\mathrm{1+3}\overbrace{-\mathrm{i}\frac{\partial}{\partial K_l}\hat{e}_m\partial_l\partial_\nu(\mathcal{P})^{(0)}A^{(0)}}^2\Big)\\
    &+\mathrm{i}\frac{1}{2}\Psi_{\mu\nu}\hat{e}_m\mathcal{P}A^{(0)}+\overbrace{\frac{1}{2}\partial_\mu\partial_\nu (\mathcal{P})^{(0)}\hat{e}_mA^{(0)}}^2\Big)\Bigg)\\
    &=\hat{e}^*_iw_k\partial_kD_{ij}\Big(\mathcal{P}(\alpha^{(1)}_j)_\mathrm{original}\overbrace{-\mathrm{i}\frac{\partial}{\partial K_l}\hat{e}_j\partial_l(\mathcal{P})^{(0)}A^{(0)}}^3\Big)+\frac{1}{2}\hat{e}^*_iw_kw_l\partial_k\partial_lD_{ij}\hat{e}_j\mathcal{P}A^{(0)}
\end{split}
\end{equation}
All the terms without the overbraces do not involve derivatives of the beam profile. Contrariwise, the terms with overbraces involve various derivatives of the profile function, and have numbers tacked to them for simplification purposes. We will first focus on simplifying the derivatives of our profile function. We will utilise the following identities to simplify our expressions:
\begin{equation}\label{eqn:K_derivatives_of_D}
    \begin{split}
        \frac{\partial}{\partial K_\mu} \frac{\partial}{\partial K_\nu} D_{im}&=(\delta_{il}\delta_{jm}-\delta_{im}\delta_{jl})(\delta_{j\mu}\delta_{l\nu}+\delta_{j\nu}\delta_{l\mu}),\\
    K_{\mu}\frac{\partial}{\partial K_\mu} \frac{\partial}{\partial K_\nu} D_{im}&= \frac{\partial}{\partial K_\nu} D_{ij}\\
    \end{split}
\end{equation}
The numbers encode various groups of terms that simplify into much simpler expressions by utilising derivatives of \(D_{ij}\). The groups(excluding (4)) simplify to the following:
\begin{equation}\label{eqn:group1}
    (1): \mathrm{i}\hat{e}^*_i\bigg(\frac{\partial}{\partial K_\mu}\frac{\partial}{\partial K_\nu}(D_{ij}\hat{e}_j)\bigg)w_k\Psi_{k\mu }\partial_\nu(\mathcal{P})^{(0)}A^{(0)}
\end{equation}
\begin{equation}\label{eqn:group2}
    (2): \hat{e}^*_i\bigg(\frac{\partial}{\partial K_\mu}\frac{\partial}{\partial K_\nu}(D_{ij}\hat{e}_j)\bigg)\frac{1}{2}\partial_\mu\partial_\nu(\mathcal{P})^{(0)}A^{(0)}
\end{equation}
\begin{equation}\label{eqn:group3}
    (3): \mathrm{i}\hat{e}^*_i\bigg(\partial_\mu\frac{\partial}{\partial K_\nu}(D_{ij}\hat{e}_j)\bigg)w_\mu\partial_\nu(\mathcal{P})^{(0)}A^{(0)}
\end{equation}
Deriving these expressions is rather straightforward, and we will outline the terms and steps we used to get each group below. Firstly, for group (1), the terms we use are:
\begin{equation}
\begin{split}    \mathrm{i}\hat{e}^*_i\bigg(&\frac{\partial}{\partial K_\nu} D_{im}\Psi_{\nu k}w_k\frac{\partial}{\partial K_l}\hat{e}_m\partial_l(\mathcal{P})^{(0)}+\frac{\partial}{\partial K_\mu}\frac{\partial}{\partial K_\nu} D_{im}\Psi_{\mu k}w_k\hat{e}_m\partial_\nu(\mathcal{P})^{(0)}\\
&+\frac{\partial}{\partial K_\nu} D_{im}\Psi_{l k}w_k\frac{\partial}{\partial K_l}\hat{e}_m\partial_\nu(\mathcal{P})^{(0)}\bigg)A^{(0)}.
\end{split}
\end{equation}
By relabelling dummy indices, it is clear how we simplified it to the form in eq. \ref{eqn:group1}.\\
\\
For group (2), the terms we used are:
\begin{equation}
    \frac{1}{2}\hat{e}^*_i\left(\frac{\partial}{\partial K_\mu}\frac{\partial}{\partial K_\nu} D_{im}\partial_\mu\partial_\nu (\mathcal{P})^{(0)}\hat{e}_m+2\frac{\partial}{\partial K_\nu} D_{im}\partial_l\partial_\nu (\mathcal{P})^{(0)}\frac{\partial}{\partial K_l}\hat{e}_m\right)A^{(0)}.
\end{equation}
By using the symmetry of the Hessian operator and once again relabelling dummy indices, we arrive at the expression in eq. \ref{eqn:group2}.\\
\\
For group (3), the terms we used are:
\begin{equation}
    \mathrm{i}\hat{e}^*_i\left(\frac{\partial}{\partial K_\nu} D_{im}w_k\partial_k\hat{e}_m\partial_\nu(\mathcal{P})^{(0)}+\partial_k D_{im}w_k\frac{\partial}{\partial K_l}\hat{e}_m\partial_l(\mathcal{P})^{(0)}\right)A^{(0)}.
\end{equation}
Since \(\partial_\mu\frac{\partial}{\partial K_\nu} D_{ij}=0\), relabelling dummy indices gets us the expression in eq. \ref{eqn:group3}. Thus, all the terms involving derivatives of the profile function add up to the following expression:
\begin{equation}\label{eqn:profile-derivatives-simplified}
\begin{split}
    \mathrm{i}&\Big(\partial_\tau(\mathcal{P})+w_y\frac{\mathrm{d}\hat{\bm{x}}}{\mathrm{d}\tau}\cdot\hat{\bm{y}}\partial_{w_x}(\mathcal{P})+w_x\frac{\mathrm{d}\hat{\bm{y}}}{\mathrm{d}\tau}\cdot\hat{\bm{x}}\partial_{w_y}(\mathcal{P})\Big)A^{(0)}+\mathrm{i}\hat{e}^*_i\left(\frac{\partial}{\partial K_\mu}\frac{\partial}{\partial K_\nu}(D_{ij}\hat{e}_j)\right)w_k\Psi_{k\mu }\partial_\nu(\mathcal{P})^{(0)}A^{(0)}\\
    &+\hat{e}^*_i\left(\frac{\partial}{\partial K_\mu}\frac{\partial}{\partial K_\nu}(D_{ij}\hat{e}_j)\right)\frac{1}{2}\partial_\mu\partial_\nu(\mathcal{P})^{(0)}A^{(0)}+\mathrm{i}\hat{e}^*_i\left(\partial_\mu\frac{\partial}{\partial K_\nu}(D_{ij}\hat{e}_j)\right)w_\mu\partial_\nu(\mathcal{P})^{(0)}A^{(0)}.
\end{split}
\end{equation}
At this juncture, we are free to make some definitions. We will define all the terms proportional to \(\mathcal{P}\) to be 0, this is ultimately an arbitrary choice we make for simplicity and to maintain compatibility with previous work, but mathematically this pushes all evolution dependence to \(\mathcal{P}\). Mathematically, this is akin to changing a differential equation in \(f(x)=g(x)h(x)\) into one in \(g(x)\) simply by specifying a specific form for \(h(x)\). \\
\\
In order for the terms proportional to \(\mathcal{P}\) to be 0, we see that there are terms proportional to \(\bm{w}\bm{w}\) and those independent of it. They therefore have to be separately 0.\\
\\
In the subsequent section deriving the beam envelope evolution and the amplitude function evolution, all instances of \(\frac{\mathrm{d}}{\mathrm{d}\tau}\) acting on index notation components should be taken as covariant derivatives, that is:
\begin{equation}
    \frac{\mathrm{d}}{\mathrm{d}\tau}(T_{\mu_0\mu_1\mu_2...\mu_n}):=\frac{\partial T_{\mu_0\mu_1\mu_2...\mu_n}}{\partial\tau}+\sum_{i=0}^n\sum_{j=0}^2\hat{x}_j^{\mu_n}\frac{\mathrm{d}\hat{x}_j^\rho}{\mathrm{d}\tau}T_{\mu_0...\mu_{n-1}\rho\mu_{n+1}...\mu_n}.
\end{equation}
Note that \((\hat{\bm{x}}_0,\hat{\bm{x}}_1,\hat{\bm{x}}_2):=(\hat{\bm{g}},\hat{\bm{x}},\hat{\bm{y}})\) in our analysis. Intuitively, we are saying that when you see the \(\frac{\mathrm{d}}{\mathrm{d}\tau}\) operator at any subsequent points in this manuscript, you can take it to refer to the covariant operator that takes into account the change in the basis directions. When we later use tensor notation, this covariant operator is not needed as the \(\tau\) derivative operates on the mathematical objects themselves, using the covariant differentiation inherited from the embedding into Euclidean space.
\subsubsection{Terms proportional to \(\bm{w}\bm{w}\)}
From eq.\ref{eqn:second_order_helmholtz}, we can see that the terms we wish to set to 0 are:
\begin{equation}
\begin{split}
    &-\frac{1}{2}w_kw_\rho\Bigg(\hat{e}^*_i\frac{\partial}{\partial K_\nu}(D_{im})\hat{e}_m\frac{\hat{g}_\nu}{g}\left(\frac{\mathrm{d}\Psi_{w,k\rho}}{\mathrm{d}\tau}+2\frac{\mathrm{d}\hat{g}_{k}}{\mathrm{d}\tau}\frac{\mathrm{d}K_{\rho}}{\mathrm{d}\tau}\right)\\
    &\qquad+\Psi_{k\mu}\Psi_{\rho\nu}\hat{e}^*_i\frac{\partial}{\partial K_\mu}\frac{\partial}{\partial K_\nu}(D_{im})\hat{e}_m+2\hat{e}^*_i\partial_k(D_{im})(\Psi_{\rho l}\frac{\partial}{\partial K_l}+\partial_\rho)\hat{e}_m\\
    &\qquad+\hat{e}^*_i\partial_\mu\partial_\nu(D_{im})\hat{e}_m+2\hat{e}^*_i\Psi_{k\nu}\frac{\partial}{\partial K_\nu}(D_{im})(\Psi_{\rho l}\frac{\partial}{\partial K_l}+\partial_\rho)\hat{e}_m\Bigg)=0.
\end{split}
\end{equation}
Only the symmetric part remains as the coefficient, thus we just have to symmetrise the tensor contracted by \(w_kw_\rho\). To simplify our resulting expression, we will substitute in the ray-tracing equations. We will utilise the first equality of the following identity:
\begin{equation}\label{eqn:useful_Hamiltonian_trick}
\begin{split}
\hat{e}^*_i\partial_\mu\partial_\nu(D_{im}\hat{e}_m)&=\hat{e}^*_i\partial_\mu\partial_\nu(D_{im})\hat{e}_m+\hat{e}^*_i\partial_\mu D_{im}\partial_\nu\hat{e}_m+\hat{e}^*_i\partial_\nu D_{im}\partial_\mu\hat{e}_m,\\
&=\partial_\mu\partial_\nu H+\partial_\mu H \hat{e}^*_i\partial_\nu\hat{e}_i+\partial_\nu H \hat{e}^*_i\partial_\mu\hat{e}_i.\\
\end{split}
\end{equation}
Note that anything along \(\bm{g}\) is contracted to 0 by \(w_kw_\rho\). We can utilise this throughout the expression, but more usefully, we can notice that:
\begin{equation}
    \left(\frac{\mathrm{d}\Psi_{w,\mu\nu}}{\mathrm{d}\tau}+\frac{\mathrm{d}\hat{g}_{\mu}}{\mathrm{d}\tau}\frac{\mathrm{d}K_{\nu}}{\mathrm{d}\tau}+\frac{\mathrm{d}\hat{g}_{\nu}}{\mathrm{d}\tau}\frac{\mathrm{d}K_{\mu}}{\mathrm{d}\tau}\right)_w=\left(\frac{\mathrm{d}\Psi_{k\rho}}{\mathrm{d}\tau}\right)_w.
\end{equation}
This yields the following equation:
\begin{equation}
\begin{split}
    &\frac{1}{2}w_kw_\rho\Bigg(\frac{\mathrm{d}\Psi_{k\rho}}{\mathrm{d}\tau}+\Psi_{k\mu}\Psi_{\rho\nu}\hat{e}^*_i\frac{\partial}{\partial K_\mu}\frac{\partial}{\partial K_\nu}(D_{im}\hat{e}_m)+\hat{e}^*_i\partial_k\partial_\rho(D_{im}\hat{e}_m)\\
    &\qquad+\hat{e}^*_i\Psi_{k\nu}\frac{\partial}{\partial K_\nu}\partial_\rho(D_{im}\hat{e}_m)+\hat{e}^*_i\Psi_{\rho\nu}\frac{\partial}{\partial K_\nu}\partial_k(D_{im}\hat{e}_m)\Bigg)=0.
\end{split}
\end{equation}
We then apply the second equality of eq.\ref{eqn:useful_Hamiltonian_trick}, and then call upon the fact that \(\Psi_{\mu\nu}\frac{\partial}{\partial K_\nu} H=-\partial_\mu H\) to simplify the above expression, obtaining: 
\begin{equation}\label{eqn:Beam_Tracing_ww}
\begin{split}
    \frac{1}{2}w_kw_\rho\left(\frac{\mathrm{d}\Psi_{k\rho}}{\mathrm{d}\tau}+\Psi_{k\mu}\Psi_{\rho\nu}\frac{\partial}{\partial K_\mu}\frac{\partial}{\partial K_\nu}(H)+\partial_k\partial_\rho(H)+\Psi_{k\nu}\frac{\partial}{\partial K_\nu}\partial_\rho(H)+\Psi_{\rho\nu}\frac{\partial}{\partial K_\nu}\partial_k(H)\right)=0.
\end{split}
\end{equation}
We can also differentiate \(\Psi_{\mu\nu}\frac{\partial}{\partial K_\nu} H=-\partial_\mu H\) with respect to \(\tau\) to yield the components of the beam tracing equation with a \(\bm{g}\) component. 
\begin{equation}
\begin{split}
    \frac{\mathrm{d}}{\mathrm{d}\tau}(\Psi_{\mu\nu}\frac{\partial}{\partial K_\nu} H+\partial_\mu H)&=0,\\
    g_k\left(\frac{\mathrm{d}\Psi_{k\rho}}{\mathrm{d}\tau}+\Psi_{k\mu}\Psi_{\rho\nu}\frac{\partial}{\partial K_\mu}\frac{\partial}{\partial K_\nu}(H)+\partial_k\partial_\rho(H)+\Psi_{k\nu}\frac{\partial}{\partial K_\nu}\partial_\rho(H)+\Psi_{\rho\nu}\frac{\partial}{\partial K_\nu}\partial_k(H)\right)&=0.    
\end{split}
\end{equation}
Overall then, we combine this with eq.\ref{eqn:Beam_Tracing_ww} to conclude that the following must be true:
\begin{equation}\label{eqn:Beam_Tracing_appendix}
\begin{split}
    \frac{\mathrm{d}\Psi_{k\rho}}{\mathrm{d}\tau}+\Psi_{k\mu}\Psi_{\rho\nu}\frac{\partial}{\partial K_\mu}\frac{\partial}{\partial K_\nu}(H)+\partial_k\partial_\rho(H)+\Psi_{k\nu}\frac{\partial}{\partial K_\nu}\partial_\rho(H)+\Psi_{\rho\nu}\frac{\partial}{\partial K_\nu}\partial_k(H)=0.
\end{split}
\end{equation}
These equations provide us with the evolution of the Gaussian envelope as it propagates through the plasma. More precisely, we have chosen our tensor, \(\boldsymbol{\Psi}\), to satisfy eq.\ref{eqn:Beam_Tracing}, as such a tensor trivially satisfies eq. \ref{eqn:Beam_Tracing_appendix}.
\subsubsection{Terms independent of \(\bm{w}\)}
The remaining terms(with some simple manipulation and relabelling of dummy-indices) are:
\begin{equation}
\begin{split}
    A^{(0)}\Bigg(&\frac{\mathrm{dln}(A^{(0)})}{\mathrm{d}\tau}+\frac{\hat{g}_\mu}{g}\frac{\partial}{\partial K_\mu} D_{im}\hat{e}^*_i\frac{\mathrm{d}\hat{e}_m}{\mathrm{d}\tau}+\hat{e}^*_i\frac{\partial}{\partial K_\mu} D_{im}\left(\Psi_{\mu\nu}\frac{\partial}{\partial K_\nu}+\partial_\mu\right)\hat{e}_m\\
    &-\hat{g}_k\partial K_kD_{im}\hat{e}^*_i\hat{g}_\mu\left(\Psi_{\mu\nu}\frac{\partial}{\partial K_\nu}+\partial_\mu\right)\hat{e}_m+\frac{1}{2}\Psi_{\mu\nu}\hat{e}^*_i\frac{\partial}{\partial K_\mu}\frac{\partial}{\partial K_\nu} D_{im}\hat{e}_m\Bigg)=0.
\end{split}
\end{equation}
Using the fact that \(g_\mu\Psi_{\mu\nu}=\frac{\mathrm{d}K_\nu}{\mathrm{d}\tau}\), and that \(g_\mu\partial_\mu+\frac{\mathrm{d}K_\mu}{\mathrm{d}\tau}\frac{\partial}{\partial K_\mu}=\frac{\mathrm{d}}{\mathrm{d}\tau}\), we can then simplify into the following equation:
\begin{equation}\label{eqn:second_order_terms_no_w}
\begin{split}
    \frac{\mathrm{dln}(A^{(0)})}{\mathrm{d}\tau}+\frac{\mathrm{d}K_\mu}{\mathrm{d}\tau}\hat{e}^*_m\frac{\partial}{\partial K_\mu} \hat{e}_m+\hat{e}^*_i\frac{\partial}{\partial K_\mu} D_{im}\partial_\mu\hat{e}_m+\frac{1}{2}\Psi_{\mu\nu}\frac{\partial}{\partial K_\mu}\frac{\partial}{\partial K_\nu} H=0.
\end{split}
\end{equation}
With this, we then decompose the amplitude into its real and imaginary parts, since \(\ln(A^{(0)})=\ln|A^{(0)}|+\mathrm{i}\phi\) and analyse each separately. Using the fact that:
\begin{equation}
    \partial_\mu\frac{\partial}{\partial K_\mu}(\hat{e}^*_iD_{im}\hat{e}_m)=\hat{e}^*_i\frac{\partial}{\partial K_\mu} D_{im}\partial_\mu\hat{e}_m+\partial_\mu\hat{e}^*_i\frac{\partial}{\partial K_\mu} D_{im}\hat{e}_m.
\end{equation}
Thus, what we find is the following for the real part of eq. \ref{eqn:second_order_terms_no_w}:
\begin{equation}
    \frac{\mathrm{dln}(|A^{(0)}|)}{\mathrm{d}\tau}+\frac{1}{2}\frac{\partial}{\partial K_\mu}\partial_\mu H+\frac{1}{2}\Re(\Psi_{\mu\nu})\frac{\partial}{\partial K_\mu}\frac{\partial}{\partial K_\nu} H=0.
\end{equation}
Then, we utilise eq. \ref{eqn:Beam_Tracing_appendix} to obtain:
\begin{equation}
\begin{split}
    0&=\Im(\Psi)^{-1}_{\mu\nu}\Bigg(\frac{d\Im(\Psi)_{\mu\nu}}{d\tau}+\Re(\Psi)_{\mu\rho}\partial K_\rho\partial K_\sigma H\Im(\Psi)_{\sigma\nu}+\Im(\Psi)_{\mu\rho}\partial K_\rho\partial K_\sigma H\Re(\Psi)_{\sigma\nu}\\
    &+\Im(\Psi)_{\mu\rho}\partial K_\rho\partial_\nu H+\partial_\mu\partial K_\sigma H\Im(\Psi)_{\sigma\nu}\Bigg),\\
\end{split}
\end{equation}
which we use with:
\begin{equation}
    \Im(\Psi)^{-1}_{\mu\nu}\frac{d\Im(\Psi)_{\mu\nu}}{d\tau}=\frac{\mathrm{dln}\left(\mathrm{Det(}\Im(\Psi)_{\mu\nu})\right)}{\mathrm{d}\tau},
\end{equation}
along with:
\begin{equation}
    \Im(\Psi)^{-1}_{\mu k}\Im(\Psi)_{k \nu}=\delta_{\mu\nu}-\hat{g}_\mu\hat{g}_\nu,
\end{equation}
to arrive at:
\begin{equation}
    \frac{\mathrm{dln}(|A^{(0)}|)}{\mathrm{d}\tau}-\frac{1}{4}\left(\frac{\mathrm{dln}\left(\mathrm{Det(}\Im(\Psi)_{\mu\nu})\right)}{\mathrm{d}\tau}-2\hat{g}_\mu\hat{g}_\nu(\frac{\partial}{\partial K_\mu}\partial_\nu H+\Re(\Psi_{\mu k})\partial K_k\frac{\partial}{\partial K_\nu} H)\right)=0.
\end{equation}
We now use \(g_\mu\Re(\Psi_{\mu\nu})=\frac{\mathrm{d}K_\nu}{\mathrm{d}\tau}\) to obtain:
\begin{equation}
    \frac{\mathrm{dln}(|A^{(0)}|)}{\mathrm{d}\tau}=\frac{1}{4}\frac{\mathrm{dln}\left(\mathrm{Det(}\Im(\Psi)_{\mu\nu})\right)}{\mathrm{d}\tau}-\frac{1}{2g}\frac{\mathrm{d}g}{\mathrm{d}\tau},
\end{equation}
from which we simply get that:
\begin{equation}\label{eqn:Amplitude_Evolution_magnitude_appendix}
    |A^{(0)}|=C\frac{\mathrm{Det(}\Im(\Psi)_{\mu\nu})^\frac{1}{4}}{g^\frac{1}{2}}.
\end{equation}
Next, we take a look at the imaginary part:
\begin{equation}
\begin{split}
    &\frac{\mathrm{d}\phi}{\mathrm{d}\tau}+\frac{1}{\mathrm{i}}\frac{\mathrm{d}K_\mu}{\mathrm{d}\tau}\hat{e}^*_m\frac{\partial}{\partial K_\mu} \hat{e}_m+\frac{1}{2}\Im(\Psi_{\mu\nu})\frac{\partial}{\partial K_\mu}\frac{\partial}{\partial K_\nu} H\\
    &+\frac{1}{2\mathrm{i}}\left(\hat{e}^*_i\frac{\partial}{\partial K_\mu} D_{im}\partial_\mu\hat{e}_m-\partial_\mu\hat{e}^*_i\frac{\partial}{\partial K_\mu} D_{im}\hat{e}_m\right)=0.
\end{split}
\end{equation}
From here we just use the fact that \(\frac{\partial}{\partial K_\mu} H\hat{e}_i=\frac{\partial}{\partial K_\mu} D_{im}\hat{e}_m+D_{im}\frac{\partial}{\partial K_\mu}\hat{e}_m\) along with its conjugate equation to arrive at this final simple expression for the phase evolution:
\begin{equation}\label{eqn:Amplitude_Evolution_Phase_appendix}
    \frac{\mathrm{d}\phi}{\mathrm{d}\tau}=\mathrm{i}\hat{e}^*_i\frac{\mathrm{d}\hat{e}_i}{\mathrm{d}\tau}+\frac{\mathrm{i}}{2}\left(\partial_\mu\hat{e}^*_i D_{im}\frac{\partial}{\partial K_\mu}\hat{e}_m-\frac{\partial}{\partial K_\mu}\hat{e}^*_i D_{im}\partial_\mu\hat{e}_m\right)-\frac{1}{2}\Im(\Psi_{\mu\nu})\frac{\partial}{\partial K_\mu}\frac{\partial}{\partial K_\nu} H.
\end{equation}
We have two contributions to our phase, called the polarisation and Gouy phase. The polarisation phase is some change to our phase due to our changing polarisation, and is given by:
\begin{equation}
    \frac{\mathrm{d}\phi_P}{\mathrm{d}\tau}=\mathrm{i}\hat{e}^*_i\frac{\mathrm{d}\hat{e}_i}{\mathrm{d}\tau}+\frac{\mathrm{i}}{2}(\partial_\mu\hat{e}^*_i D_{im}\frac{\partial}{\partial K_\mu}\hat{e}_m-\frac{\partial}{\partial K_\mu}\hat{e}^*_i D_{im}\partial_\mu\hat{e}_m).
\end{equation}
The Gouy phase is given by:
\begin{equation}
    \frac{\mathrm{d}\phi_G}{\mathrm{d}\tau}=-\frac{1}{2}\Im(\Psi_{\mu\nu})\frac{\partial}{\partial K_\mu}\frac{\partial}{\partial K_\nu} H.
\end{equation}

We have thus successfully defined \(A^{(0)}(\tau)\) and \(\Psi_{\mu\nu}(\tau)\) such that our terms that do not involve derivatives of our profile function are set to 0.

\subsubsection{Profile Evolution Equation}
The terms left over thus specify how our profile evolves as our beam propagates through the plasma. That is, eq. \ref{eqn:profile-derivatives-simplified} must equal 0 due to our choice. 
\begin{equation}
\begin{split}
    &\mathrm{i}\frac{\partial\mathcal{P}}{\partial\tau}+\left(\frac{\partial}{\partial K_\mu}\frac{\partial}{\partial K_\nu} H\right)\frac{1}{2}\partial_\mu\partial_\nu(\mathcal{P})^{(0)}\\
    &+\mathrm{i}\Big(w_\mu\partial_\nu \mathcal{P}^{(0)}\Big)\Bigg(\left(\frac{\mathrm{d}\hat{\bm{y}}}{\mathrm{d}\tau}\cdot\hat{\bm{x}}\right)(\hat{x}_\mu\hat{y}_\nu-\hat{y}_\mu\hat{x}_\nu)\\
    &\qquad+\Big(\Psi_{\mu\rho}\partial K_\rho\frac{\partial}{\partial K_\nu} H+\partial_\mu\frac{\partial}{\partial K_\nu} H\Big)\Bigg)=0.\\
\end{split}
\end{equation}
Defining the following tensor:
\begin{equation}
\begin{split}
    T_{\mu\nu}(\tau)&=\left(\frac{\mathrm{d}\hat{\bm{y}}}{\mathrm{d}\tau}\cdot\hat{\bm{x}}\right)(\hat{x}_\mu\hat{y}_\nu-\hat{y}_\mu\hat{x}_\nu)+\Big(\Psi_{\mu\rho}\partial K_\rho\frac{\partial}{\partial K_\nu} H+\partial_\mu\frac{\partial}{\partial K_\nu} H\Big)
\end{split},
\end{equation}
We then arrive at the following simplified, key expression for the evolution of our beam profile:
\begin{equation}\label{eqn:Profile_Evolution_appendix}
    \frac{\partial\mathcal{P}}{\partial\tau}-\mathrm{i}\left(\frac{1}{2}\frac{\partial}{\partial K_\mu} \frac{\partial}{\partial K_\nu} H\partial_\mu\partial_\nu(\mathcal{P})^{(0)}\right)+w_\mu T_{\mu\nu}\partial_\nu \mathcal{P}^{(0)}=0.
\end{equation}
This is the profile evolution equation, which is a second-order 2D linear PDE. Solving for eigenprofiles of this PDE is an important task that we undertake in section \ref{Solving_Profile_Evolution}. By obtaining these eigenprofiles, we will be able to decompose any beam profile into a linear superposition of eigenprofiles, for which we know the exact evolution. This avoids the need to solve any PDEs. \\
\\
To summarise, after converting eqs. \ref{eqn:dispersion relation_appendix}, \ref{eqn:ray-tracing_appendix}, \ref{eqn:Beam_Tracing_appendix}, \ref{eqn:Amplitude_Evolution_magnitude_appendix}, \ref{eqn:Amplitude_Evolution_Phase_appendix}, \ref{eqn:Profile_Evolution_appendix} back into tensor notation, we obtain the set of equations we present in section \ref{beam-tracing-eqns}.

\section{Properties of ladder operators}\label{ladder_properties}
The ladder operators presented here are useful in a general context. They allow us to construct infinite families of solutions to this specific case of the Fokker-Planck equation\cite{Risken1996}, which bears striking resemblance to the PDE for the multidimensonal Ornstein-Uhlenbeck process\cite{Vatiwutipong2019}. They are thus useful mathematical tools that can be applied in a wider mathematical context for solving similar PDEs in any number of dimensions.
\subsection{Uniqueness of Solutions}
A simple property to demonstrate is uniqueness. Suppose that \(\hat{\mathcal{D}}\) had two possible solutions, \(f_1\) and \(f_2\). Then, \(f_3=f_1-f_2\) must be a solution. If we choose \(f_1\) and \(f_2\) such that they have the same boundary values, then \(f_3\) has a boundary value of 0. Thus, \(f_3=0\) throughout the evolution. This means that \(\hat{\mathcal{D}}\) admits unique solutions, which is a fact we will make use of when exploring properties of our ladder operators.
\subsection{Commutativity}
Before investigating the commutativity of the ladder operators, we first have to establish a useful corollary. It turns out that \(\bm{u}_{I}(\tau)\cdot\bm{v}_{J}(\tau)=\bm{u}_{I}(0)\cdot\bm{v}_{J}(0)\). We can prove this by direct differentiation.
\begin{equation}
    \frac{\mathrm{d}}{\mathrm{d}\tau}(u^\mu_Iv^\mu_J)=u^\mu_IT_{\mu\nu}v^\nu_J-u^\mu_IT_{\mu\nu}v^\nu_J=0.
\end{equation}

Investigating the commutativity of our ladder operators is crucial for deriving further properties of our ladder operators. For the sake of brevity later on, we will define:
\begin{equation}
    [\hat{\mathcal{L}}_I\,\mathrm{or}\,\hat{\mathcal{R}}_I, \hat{\mathcal{L}}_J\,\mathrm{or}\,\hat{\mathcal{R}}_J]=[\mathrm{Comm}]
\end{equation}
We can utilise uniqueness to identify a crucial property of our commutator. If \(f\) satisfies \(\hat{\mathcal{D}}f=0\), by acting the commutator on \(f\), we see that 
\begin{equation}
    [\mathrm{Comm}]\hat{\mathcal{D}}f=\hat{\mathcal{D}}[\mathrm{Comm}]f.
\end{equation}
If \([\mathrm{Comm}]f=Af\) at the boundary, then it is clear that \([\mathrm{Comm}]f=Af\) throughout, given that \textit{f} is a unique solution. Thus, the commutator of any two ladder operators must remain constant, regardless of what point during the evolution we are at. We thus only need to evaluate the commutator at a single point, and we will know what it is at all other points. This feature is built into our ladder operators, as when we explicitly evaluate the commutators, we find that:
\begin{equation}
\begin{split}
    [\hat{\mathcal{L}}_I, \hat{\mathcal{R}}_J]&=\bm{u}_{I}(\tau)\cdot\bm{v}_{J}(\tau),\\
    &=\bm{u}_{I}(0)\cdot\bm{v}_{J}(0),\\
    [\hat{\mathcal{R}}_I, \hat{\mathcal{L}}_J]&=-\bm{u}_{J}(0)\cdot\bm{v}_{I}(0),\\
    [\hat{\mathcal{L}}_I, \hat{\mathcal{L}}_J]&=[\hat{\mathcal{R}}_I, \hat{\mathcal{R}}_J]=0.
\end{split}
\end{equation}
Due to this, if \(f_0\) is the stationary state of \(\hat{\mathcal{R}}_I\), by repeatedly commuting our operators, one can show that:
\begin{equation}
    \hat{\mathcal{R}}_J \hat{\mathcal{L}}_I^n  f_0=n(-\bm{u}_{J}(0)\cdot\bm{v}_{I}(0))\hat{\mathcal{L}}_I^{n-1}  f_0,
\end{equation}
as one would expect from the uniqueness of our solutions. We can trivially perform the same process on the stationary state of \(\hat{\mathcal{L}}_I\) to obtain a similar expression. 

\subsection{Biorthogonality\label{biorthogonality}}
It turns out that in special cases, the ladder operators can exhibit biorthogonality. We will subsequently assume that we have chosen a biorthonormal basis at the boundary. We are interested in whether such a basis stays biorthogonal when subject to evolution. We will utilise \(\bm{x}=\sqrt{\mathrm{i}\bm{\Phi}^\mathrm{(Case\,1)}}\cdot\bm{w}\) to define our coordinate system in both cases. It is important to keep track of the fact that this then means that our ladder operators have different forms depending on which case we are applying them to. 
For case 1, they are:
\begin{equation}\label{eqn:ladder_case_1_definition}
\begin{split}
    \hat{\mathcal{L}}_I&=\bm{\mu}_I(\tau)\cdot\bm{\nabla}_x,\\
    \hat{\mathcal{R}}_I&=\bm{\nu}_I(\tau)\cdot(\bm{x}-\bm{\nabla}_x).
\end{split}
\end{equation}
Here, \(\bm{\mu}_I(\tau)=\boldsymbol{u}_I(\tau)\cdot\sqrt{\mathrm{i}\boldsymbol{\Phi}}\) and \(\bm{\nu}_I(\tau)=\boldsymbol{v}_I(\tau)\cdot\sqrt{\mathrm{i}\boldsymbol{\Phi}}^{-1}\) For case 2 they are:
\begin{equation}
\begin{split}
    \hat{\mathcal{L}}_I&=-\bm{\nu}_I(\tau)\cdot\bm{\nabla}_x,\\
    \hat{\mathcal{R}}_I&=\bm{\mu}_I(\tau)\cdot(\bm{x}+\bm{\nabla}_x).
\end{split}
\end{equation}
These definitions we used above are guaranteed to be true as they are the only way that the ladder operators can produce the same results in each case, asserted via uniqueness. In this coordinate system, the ground states we use to generate our solutions in each case are given by 1 and \(\exp\left(-\frac{1}{2}\bm{x}\cdot\bm{x}\right)\) respectively. Note that when we calculated the commutators, we demonstrated earlier that \(\bm{u}_{I}\cdot\bm{v}_{J}=\bm{u}_{I}(0)\cdot\bm{v}_{J}(0)\). It then follows that:
\begin{equation}
    \bm{\mu}_{I}\cdot\bm{\nu}_{J}=\bm{\mu}_{I}(0)\cdot\bm{\nu}_{J}(0).
\end{equation}
If $\bm{\nu}_{J}(\tau)$ remain orthogonal, we then have that:
\begin{equation}
    \bm{\mu}_{I}(\tau)=\sum_J\frac{\bm{\mu}_{I}(0)\cdot\bm{\nu}_{J}(0)}{|\bm{\nu}_{J}(\tau)|^2} \bm{\nu}_{J}(\tau).
\end{equation}
In the subsequent section, we will assume that $\bm{\nu}_{J}(\tau)$ are orthogonal, and use this to investigate the inner-product of any two states of our system. We will assume that our space is N-dimensional, and that \(\bm{u}_{I}(0)\cdot\bm{v}_{J}(0)=\delta_{IJ}\). We keep the case 1 state(\(g_{n_1...n_i...n_N}\)) on the left and the case 2 state(\(f_{m_1...m_i...m_N}\)) on the right, and define the following inner product:
\begin{equation}
    \langle g_{n_1...n_i...n_N},f_{m_1...m_i...m_N}\rangle=\int_{-\infty}^\infty g_{n_1...n_i...n_N}f_{m_1...m_i...m_N}\mathrm{d}^Nx.
\end{equation}
Assuming that \(g_{n_1...n_i...n_N}f_{m_1...m_i...m_N}\) goes to 0 at \(\pm\infty\), and since the ladder operators commute without changing our function(beyond scaling), we then find that:
\begin{equation}\label{eqn:ladder_case}
    \langle g_{n_1...n_i...n_N},\hat{\mathcal{L}}_i^{k}(\mathrm{Case\,2})f_{m_1...m_i-k...m_N}\rangle=|\bm{\nu}_i(\tau)|^2\langle \hat{\mathcal{L}}_i^{k}(\mathrm{Case\,1})g_{n_1...n_i...n_N},f_{m_1...m_i-k...m_N}\rangle,
\end{equation}
and 
\begin{equation}\label{eqn:ladder_case_adjoint}
    |\bm{\nu}_i(\tau)|^2\langle g_{n_1...n_i...n_N},\hat{\mathcal{R}}_i^k(\mathrm{Case\,2})f_{m_1...m_i+k...m_N}\rangle=\langle \hat{\mathcal{R}}_i^k(\mathrm{Case\,1})g_{n_1...n_i...n_N},f_{m_1...m_i+k...m_N}\rangle.
\end{equation}
These equations tell us that we can move ladder operators between our case 1 and case 2 functions within the inner product. It then follows that if there is any \(n_i<m_i\) or \(n_i>m_i\), we can keep moving ladder operators from one function to another until we get 0 for one of the functions, since the functions are generated from stationary states of the ladder operators. Thus, unless \((n_1...n_i...n_N)=(m_1...m_i...m_N)\), our inner product will yield 0. \\
\\
To summarise the results of this section, we have found that if the ladder operators produce a biorthogonal basis of solutions at the boundary, our solutions will remain biorthogonal throughout the evolution. We will define the following inner product:
\begin{equation}\label{eqn:gaussian_inner_product}
    \langle f_{n_1...n_i...n_N},f_{m_1...m_i...m_N}\rangle=\int_{-\infty}^\infty f_{n_1...n_i...n_N}f_{m_1...m_i...m_N}\exp\left(-\frac{1}{2}\bm{x}\cdot\bm{x}\right)\mathrm{d}^Nx,
\end{equation}
where \(f_{n_1...n_i...n_N}\) and \(f_{m_1...m_i...m_N}\) are functions generated by repeated application of the ladder operators in case 1. What we have proven is that under this inner product, \(\langle f_{n_1...n_i...n_N},f_{m_1...m_i...m_N}\rangle=\delta_{(n_1...n_i...n_N)(m_1...m_i...m_N)}\). Furthermore, using the definition in eq. \ref{eqn:ladder_case_1_definition}, we have also shown from eq. \ref{eqn:ladder_case} and eq. \ref{eqn:ladder_case_adjoint} that with this inner product, 
\begin{equation}
\begin{split}
    \langle f_{n_1...n_i...n_N},\hat{\mathcal{L}}_i f_{m_1...m_i...m_N}\rangle&=|\bm{\nu}_i(\tau)|^2\langle \hat{\mathcal{R}}_i f_{n_1...n_i...n_N},f_{m_1...m_i...m_N}\rangle,\\
    |\bm{\nu}_i(\tau)|^2\langle f_{n_1...n_i...n_N},\hat{\mathcal{R}}_i f_{m_1...m_i...m_N}\rangle&=\langle \hat{\mathcal{L}}_i f_{n_1...n_i...n_N},f_{m_1...m_i...m_N}\rangle.
\end{split}
\end{equation}
Note that this all hinges on the assumption that $\bm{\nu}_{J}(\tau)$ are orthogonal, which is not true in general. Thus, unless $\bm{\nu}_{J}(\tau)$ are orthogonal, even though our solutions form a biorthogonal basis at the plasma boundary, we are not assured that they will remain in such a biorthogonal basis for a generic non-zero $\tau$. To summarise, although the ladder operators certainly act like inverses to each other, we are not guaranteed that they will also function as adjoint operators to each other, hence we opted to address them is left($\mathcal{L}$) and right($\mathcal{R}$) operators. In general, solutions that form a biorthogonal basis at $\tau=0$ will not remain in such a basis throughout the plasma.

\end{document}